\documentclass[a4paper,11pt]{article}
\usepackage{jheppub}

\usepackage[T1]{fontenc}
\usepackage[utf8]{inputenc}
\usepackage[english]{babel}

\usepackage{caption}
\usepackage{float}

\usepackage{booktabs}
\usepackage{array}
\usepackage{multirow} 
\usepackage{makecell}
\usepackage{longtable}
\usepackage{float}
\usepackage{placeins}

\usepackage[colorinlistoftodos,%
            prependcaption,%
            textsize=scriptsize,%
            linecolor=red,%
            backgroundcolor=red!25,%
            bordercolor=red]{todonotes} 

\makeatletter
\def\cpartline#1{\@cpartline#1\@nil}
\def\@cpartline#1-#2\@nil{%
  \omit
  \@multicnt#1%
  \advance\@multispan\m@ne
  \ifnum\@multicnt=\@ne\@firstofone{&\omit}\fi
  \@multicnt#2%
  \advance\@multicnt-#1%
  \advance\@multispan\@ne
  \kern32pt
        \leaders\hrule\@height\arrayrulewidth\hfill
  \kern32pt
  \cr
  \noalign{\vskip-\arrayrulewidth}}
\makeatother

\newcommand{\mcname}{\textsc{Helac+OpenLoops}}

\newcommand{\diagwidthfraction}{0.65}

\newcommand{\rsec}[1]{section~\ref{#1}}

\usepackage[normalem]{ulem}

\title{\boldmath Higgs interference effects in top-quark pair production in the 1HSM}

\author[a]{Andrea Banfi,}
\author[b]{Nikolas Kauer,}
\author[c]{Alexander Lind,}
\author[a]{Jonas M. Lindert,}
\author[a]{and Ryan Wood}

\affiliation[a]{\footnotesize Department of Physics and Astronomy, University of Sussex, Sussex House, Brighton, BN1 9RH, U.K.}
\affiliation[b]{\footnotesize Department of Physics, Royal Holloway, University of London, Egham Hill, Egham, TW20 0EX, U.K.}
\affiliation[c]{\footnotesize SUBATECH, Universit\'e de Nantes, IMT Atlantique, IN2P3/CNRS, 4 rue Alfred Kastler, 44307 Nantes cedex 3, France}

\emailAdd{a.banfi@sussex.ac.uk}
\emailAdd{n.kauer@rhul.ac.uk}
\emailAdd{alexander.lind@subatech.in2p3.fr}
\emailAdd{j.lindert@sussex.ac.uk}
\emailAdd{rw380@sussex.ac.uk}

\abstract{We present a next-to-leading-order (NLO) study of the process $pp \; ( \to \{ h_1,\allowbreak h_2 \}) \to t\bar{t} + X$ in the 1-Higgs-singlet extension of the Standard Model with an additional heavy Higgs boson $h_2$ that mixes with the light Higgs boson $h_1$.  This process is subject to  large interference effects between loop-induced Higgs-mediated amplitudes and the QCD continuum background which tend to overcompensate any resonance contributions. A reliable modelling of the resulting top-pair invariant mass shapes requires the inclusion of higher-order QCD corrections, which are presented here. The computation of these NLO corrections is exact in all contributions but in the class of non-factorisable two-loop diagrams which are included in an approximate way such that all infrared singular limits are preserved. We present numerical results for several benchmark points with heavy Higgs masses in the range $700$--$3000$ GeV considering the production of stable top quarks. We find that the interference effects dominate the BSM signal yielding sharp dip structures instead of resonance peaks. The significance and excludability of the BSM effect is explored for the LHC Run 2, Run 3 and HL-LHC.}


\arxivnumber{2309.16759}

\begin{document}
\maketitle
\flushbottom


\section{Introduction}
\label{sec:intro}

The discovery of a new scalar boson with a mass of $125$ GeV by the ATLAS and CMS experiments at the LHC at CERN in 2012 \cite{Aad:2012tfa, Chatrchyan:2012ufa} was a milestone in particle physics. This boson is consistent with the Higgs boson predicted by the Standard Model (SM) Higgs mechanism \cite{HIGGS1964132, PhysRevLett.13.508, PhysRev.145.1156, PhysRevLett.13.321, PhysRevLett.13.585}, which provides a gauge-invariant explanation for the origin of mass for the electroweak gauge bosons by breaking the electroweak symmetry spontaneously.

There is a multitude of well-motivated scenarios for models of physics beyond the Standard Model (BSM). The majority of these models assume that a complex $\mathrm{SU}(2)$ doublet gets a vacuum expectation value and gives rise to a physical Higgs boson. New physics scenarios, such as supersymmetry, dark matter, axions and others, feature additional Higgs-like scalar particles. Adding only an extra singlet scalar, one obtains the 1-Higgs-singlet model (1HSM). It introduces a real scalar field that is neutral under the SM gauge groups. In a next-to-minimal model an additional Higgs doublet is added to the SM particle content resulting in the class of Two Higgs Doublet Models (2HDM). In this paper we consider the 1HSM as a proof-of-concept model for extensions of the SM with additional  heavy scalars.

In the 1HSM, the doublet and singlet fields mix with each other, resulting in two physical scalar states that are linear combinations of the original fields. This mixing affects the properties and interactions of the scalar bosons, and can be constrained by the experimental data from the LHC and other experiments.
One of the key motivations for the 1HSM is that it can accommodate a first-order electroweak phase transition, which is necessary for electroweak baryogenesis \cite{Profumo:2014opa,Papaefstathiou:2020iag,Papaefstathiou:2021glr,Goncalves:2022wbp}. 
The 1HSM has been extensively investigated \cite{Binoth:1996au, Schabinger:2005ei, Patt:2006fw, Bowen:2007ia, Barger:2007im, Barger:2008jx, Bhattacharyya:2007pb, Dawson:2009yx, Bock:2010nz, Fox:2011qc, Englert:2011yb, Englert:2011us, Batell:2011pz, Englert:2011aa, Gupta:2011gd, Dolan:2012ac, Batell:2012mj, No:2013wsa, Coimbra:2013qq, Profumo:2014opa, Logan:2014ppa, Chen:2014ask, Costa:2014qga, Falkowski:2015iwa, MartinLozano:2015vtq, Kanemura:2015fra, Kanemura:2016lkz, Kanemura:2017gbi, Lewis:2017dme, Casas:2017jjg, Altenkamp:2018bcs, Abouabid:2021yvw}, and the remaining parameter space following LHC constraints has been explored in refs.~\cite{Pruna:2013bma, Robens:2015gla, Robens:2016xkb, Ilnicka:2018def,DiMicco:2019ngk} including constraints from dedicated direct searches for a heavy scalar as presented for instance in refs.~\cite{ATLAS:2021ifb,ATLAS:2022hwc}.

Finding a heavy scalar at the LHC can be very challenging, especially if its couplings to electroweak gauge bosons are suppressed, and its couplings to third generation fermions are enhanced, as is the case in many SM extensions. In such a scenario, the dominant signature arises from the
heavier scalar decaying into $t \bar t$ final states. The production of a $t \bar t$ pair via the decay of a heavy scalar features a well-known interesting complication: there are large interference effects between the  Higgs signal process and $pp \to t\bar t$ continuum QCD  production~\cite{Gaemers:1984sj,Dicus:1994bm,Frederix:2007gi,Jung:2015gta,Djouadi:2016ack,Gori:2016zto,Carena:2016npr}. These large interference effects induce significant distortions to the resonance line-shape yielding different forms: a pure resonance peak, a peak-dip structure, a dip-peak structure, or even a pure dip structure, depending on the parameters of the theory. Therefore, good control of the signal-background interference 
including higher-order corrections in perturbative QCD is mandatory for a reliable interpretation of a possible future signal, but also for setting accurate exclusion limits~\cite{
Czakon:2016vfr,Bernreuther:2015fts,Hespel:2016qaf,Bernreuther:2017yhg,BuarqueFranzosi:2017jrj,BuarqueFranzosi:2017qlm,Basler:2019nas,Djouadi:2019cbm,Kauer:2019qei,Frixione:2020hqz,Atkinson:2020uos}. 
Computing next-to-leading order (NLO) corrections to the interference is non-trivial, as the interference in question at leading order (LO) involves a loop amplitude for the signal and a tree-level amplitude for the continuum. Such a structure is not supported by general purpose NLO frameworks, and at NLO order multi-scale two-loop amplitudes contribute, whose computation is beyond current technology. In refs.~\cite{
Bernreuther:2015fts,Hespel:2016qaf,Bernreuther:2017yhg,BuarqueFranzosi:2017jrj,Kauer:2019qei}
 different approximations have been employed and investigated for the computation of approximate NLO corrections to the interference contribution.  In ref.~\cite{Hespel:2016qaf}, an NLO K-factor has been constructed as the average of the K-factors for the signal and the continuum. The computations in ref.~\cite{Bernreuther:2015fts,Bernreuther:2017yhg,BuarqueFranzosi:2017jrj} are based on the large top-quark limit, which is clearly violated for the process at hand. Ref.~\cite{Kauer:2019qei} only includes a subset of all NLO contributions. 
Here we present a new NLO computation for Higgs-mediated $t \bar t$ production. For the computation of the interference we include the exact amplitude contributions everywhere but for the non-factorisable corrections. Complete NLO corrections are included for the heavy scalar signal and the continuum.

This paper is organised as follows:  
In \rsec{sec:model} we discuss the 1HSM and specify the used benchmark points.  
In \rsec{sec:nlocorrinterf} we review the details of the structure of our NLO QCD calculation. 
In \rsec{sec:calcdetails} we present the numerical setup of the computation including the Higgs decay widths, the used one-loop amplitudes and their numerical stability and describe the used one- and two-loop form factors for gluon-fusion Higgs production.  
In \rsec{sec:resultsstable}, we present cross sections and differential distributions for the Higgs signal and its interference in the 1HSM for the process $pp \; ( \to \{ h_1,\allowbreak h_2 \}) \to t\bar{t} + X$ at NLO QCD including a discussion of theoretical uncertainties.  
We conclude in \rsec{sec:conclusion}.


\section{\boldmath The $1$-Higgs-singlet model}
\label{sec:model}

In this section we introduce the 1HSM, which for us serves as a 
minimal theoretically consistent BSM model with two physical scalar bosons.
In the 1HSM, the SM Higgs sector is extended by an additional real
scalar field, which is a singlet under all gauge groups of the
SM, and which, like the SM Higgs, acquires a vacuum expectation value
(VEV) under electroweak symmetry breaking. A detailed description of
the model can for example be found in refs.~\cite{Chen:2014ask,
  LHCHiggsCrossSectionWorkingGroup:2013rie}.

The most general gauge-invariant potential for the 1HSM can be written as \cite{Schabinger:2005ei, Bowen:2007ia}
\begin{equation}
	V = \lambda \left ( \phi^{\dagger}\phi - \frac{v^2}{2} \right )^2 + \frac{1}{2} M^2 s^2 + \lambda_1 s^4 + \lambda_2 s^2 \left ( \phi^{\dagger}\phi - \frac{v^2}{2} \right ) + \mu_1 s^3 + \mu_2 s \left ( \phi^{\dagger}\phi - \frac{v^2}{2} \right ) \,, \label{eq:generalpotentail}
\end{equation}
where $s$ is the additional real singlet scalar with explicit mass $M$, which mixes with the SM $\text{SU}(2)$ Higgs doublet $\phi$.   
In order to avoid vacuum instability, and that the potential is unbounded from below, the quartic couplings $\lambda, \lambda_1, \lambda_2$ must satisfy
\begin{equation}
	\lambda > 0 \,, \quad \lambda_1 > 0 \,, \quad \lambda_2 > -2 \sqrt{\lambda \lambda_1} \,, \label{eq:1hsmcouplingconstraints}
\end{equation}
whereas the trilinear couplings $\mu_1, \mu_2$ can be either positive or negative.

In the unitary gauge, after EW symmetry breaking, the SM Higgs doublet
$\phi$ can be written as
\begin{equation}
	\phi = \frac{1}{\sqrt{2}} \begin{pmatrix}
0 \\ 
H + v
\end{pmatrix} \,,
\end{equation}
with the VEV $v \simeq 246$ GeV.%
\footnote{Note that the freedom to shift the value of $s$, so that it does not acquire a VEV, has been used (\cite{LHCHiggsCrossSectionWorkingGroup:2013rie}, Sec.~13.3.2).}  
In this gauge, the potential in eq.~\eqref{eq:generalpotentail} can be
rewritten in terms of the SM Higgs scalar $H$ and the new singlet
scalar $s$,
\begin{equation}
	V = \frac{\lambda}{4} H^4 + \lambda v^2 H^2 + \lambda v H^3 + \frac{1}{2} M^2 s^2 + \lambda_1 s^4 + \frac{\lambda_2}{2} H^2 s^2 + \lambda_2 v H s^2 + \mu_1 s^3 + \frac{\mu_2}{2} H^2 s + \mu_2 v H s \,. 
\end{equation}
The resulting mass eigenstates can be parameterised in terms of a mixing angle $\theta$,
\begin{equation}
\begin{split}
	h_1 &= H \cos\theta - s \sin\theta \,, \\
	h_2 &= H \sin\theta + s \cos\theta \,,
\end{split}
\end{equation}
where $h_1, h_2$ constitute the physical Higgs bosons of the 1HSM extension, and
\begin{equation}
	\tan(2\theta) = \frac{-\mu_2 v}{\lambda v^2 - \frac{1}{2} M^2} \,,
\end{equation}
with
\begin{equation}
	-\frac{\pi}{4} < \theta < \frac{\pi}{4} \,,
\end{equation}
under the condition $M^2 > 2 \lambda v^2$. The model has six independent parameters, which here are chosen to be $M_{h_1}$, $M_{h_2}$, $\theta$, $\mu_1$, $\lambda_1$, and $\lambda_2$. In terms of these independent parameters, the Lagrangian parameters are given by
\begin{align}
	\lambda &= \frac{\cos(2\theta) \left ( M_{h_1}^2 - M_{h_2}^2 \right ) + M_{h_1}^2 + M_{h_2}^2}{4 v^2} \,, \label{eq:1hsmlambda} \\
	M^2 &= \frac{M_{h_2}^2 - M_{h_1}^2 + \sec(2\theta) \left ( M_{h_1}^2 + M_{h_2}^2 \right )}{2 \sec(2\theta)} \,, \\
	\mu_2 &= -\tan(2\theta) \frac{\lambda v^2 - \frac{1}{2} M^2}{v} \,.
\end{align}
The physical $h_1$ state is assumed to be the light SM-like Higgs boson with a mass of $M_{h_1} = 125\,$GeV.

The Yukawa couplings of the light and heavy Higgs bosons to the top quark are modified, and become functions of the mixing angle,
\begin{equation}
	y_{t}^{h_1} = \cos(\theta) \sqrt{2} \frac{m_t}{v} \,, \hspace{10mm} y_{t}^{h_2} = -\sin(\theta) \sqrt{2} \frac{m_t}{v} \,.
\end{equation}

In our study we consider four different benchmark masses for the heavy Higgs,
$M_{h_2} = \{ 700, \allowbreak 1000, \allowbreak 1500, \allowbreak
3000 \}\,$GeV. For each heavy Higgs mass, two different values of the
mixing angle, $\theta=\{\theta_1, \theta_2\}$, are considered. These benchmark scenarios are
summarised in table~\ref{table:1hsmmodel}. The lower values
of $\theta$ are consistent with theoretical and current experimental
constraints \cite{Robens:2016xkb, Ilnicka:2018def}. The largest mixing angle, $\theta_2=\pi/8$, is no longer compatible with experimental constraints, but is included to illustrate the dependence on the mixing angle.
The perturbativity
constraint $\vert \lambda \vert < 4 \pi$ along with
eq.~\eqref{eq:1hsmlambda} imposes the condition
$\vert \theta \vert < \theta_0$ which is satisfied for all eight
benchmark points, as $\theta_0 \geq 0.42$ for
$200 \text{ GeV} \lesssim M_{h_2} \leq 3$ TeV. No renormalisation
group running of the couplings has been taken into
account.

\begin{table}[t]
\centering
\begin{tabular}{lcccc}
\toprule
$M_{h_2}$ {[}GeV{]}         & $700$          & $1000$         & $1500$         & $3000$         \\
\midrule
\multirow{2}{*}{$\theta=\theta_1$} & $\pi/15$       & $\pi/15$       & $\pi/22$       & $\pi/45$       \\
                            & $\approx 0.21$ & $\approx 0.21$ & $\approx 0.14$ & $\approx 0.07$ \\
\multirow{2}{*}{$\theta=\theta_2$} & $\pi/8$        & $\pi/8$        & $\pi/12$       & $\pi/24$       \\
                            & $\approx 0.39$ & $\approx 0.39$ & $\approx 0.26$ & $\approx 0.13$ \\
\bottomrule
\end{tabular}
\caption{Higgs mixing angles $\theta=\{\theta_1, \theta_2\}$ in the 1HSM for different masses of the heavy Higgs $h_2$, all with $M_{h_1} = 125$ GeV and $\mu_1 = \lambda_1 = \lambda_2 = 0$. These constitute the eight different model benchmark points.}
\label{table:1hsmmodel}
\end{table}

All benchmark points listed in table~\ref{table:1hsmmodel} are considered with vanishing coupling parameters $\mu_1$, $\lambda_1$, and $\lambda_2$, with $\lambda_1 > 0$ in eq.~\eqref{eq:1hsmcouplingconstraints} treated as approximately zero. Nonetheless, the decay widths for $h_2 \to h_1 h_1$ and $h_2 \to h_1 h_1 h_1$ are non-zero, due to the $H$--$s$ mixing. The partial decay widths $\Gamma(h_2 \to n \times h_1)$ for $n = \{ 2, 3, 4 \}$ are given in appendix~\ref{app:higgsdecay}, table~\ref{table:hdecaystheta}.


\section{Structure of the NLO QCD computation}
\label{sec:nlocorrinterf}

\begin{figure}[b]
	\centering
	\includegraphics[width=\diagwidthfraction\linewidth*\real{1.2}]{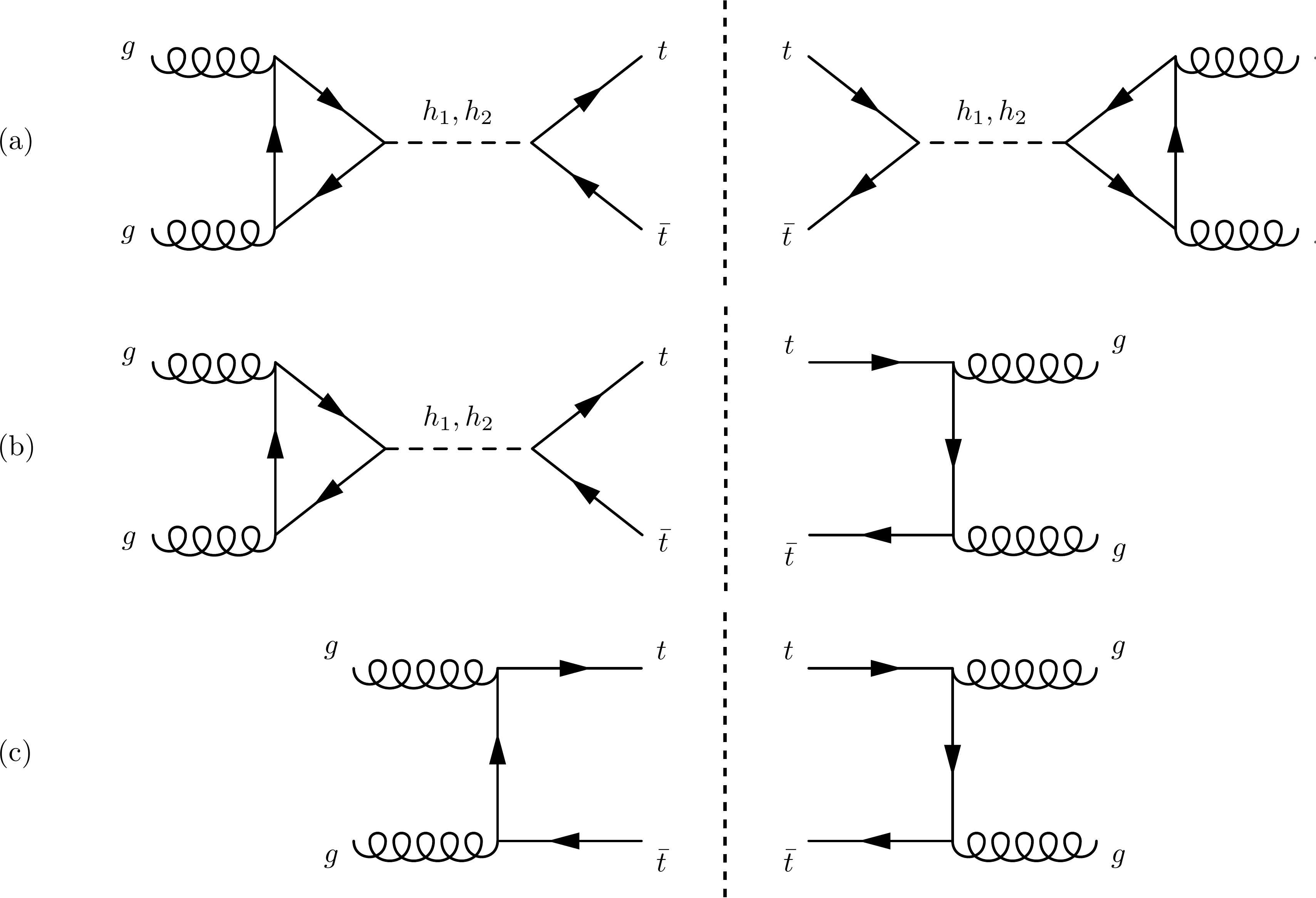}
	\caption{Leading order Feynman diagram contributions to $gg \to t \bar t$ at the amplitude-squared level: a) Higgs-squared contribution, b) Higgs-QCD interference, c) and QCD continuum-squared contribution.  The vertical dashed lines correspond to a phase-space cut on final-state particles.}
	\label{fig:lo-lo}
\end{figure}

In this paper we consider one of the key production modes of the heavy Higgs in the
1HSM extension of the SM, which is 
\begin{align}
\label{eq:process}
	gg \to \{h_1, h_2\} \to t \bar t\,. 
\end{align} 
In the SM this process is loop-induced and mediated by heavy quarks (top- and bottom-quarks). In ref.~\cite{Gaemers:1984sj,Dicus:1994bm,Frederix:2007gi,Jung:2015gta,Djouadi:2016ack,Gori:2016zto} it was pointed out that besides the squared loop-induced mode, the interference between the loop-induced signal and the $gg\to t\bar t$ background should also be taken into account for a reliable estimate of the heavy Higgs signature in the $t \bar t$ final state.
In fact, due to this interference contribution instead of a resonance peak 
one generically expects a peak-dip structure in the invariant mass distribution of the $t \bar t$ pair.
To be precise, the amplitude for the process in eq.~\eqref{eq:process} consists of the sum of QCD $t\bar t $ production, which we label $\mathcal{M}_{\text{QCD}}$, and additional contributions where the $t\bar t$ pair arises from the decay of the Higgs bosons $h_1,h_2$. We denote the corresponding amplitudes by $\mathcal{M}_{h_1}$ and $\mathcal{M}_{h_2}$. At the level of the amplitude squared, we can distinguish the following contributions,
\begin{equation}
\begin{split}
	\text{QCD background: } \; & \left \vert \mathcal{M}_{\text{QCD}} \right \vert^2 \,, \\
	\text{Higgs signal: } \; & \left \vert \mathcal{M}_{h_1} \right \vert^2 + \left \vert \mathcal{M}_{h_2} \right \vert^2 + 2 \, \mathrm{Re} \left ( \mathcal{M}_{h_1}^{\ast} \mathcal{M}_{h_2} \right ) \,, \\
	\text{Higgs--QCD interference: } \; & 2 \, \mathrm{Re} \left ( \left ( \mathcal{M}_{h_1}^{\ast} + \mathcal{M}_{h_2}^{\ast} \right ) \mathcal{M}_{\text{QCD}} \right ) \,.
\label{eq:1hsmcontributions}
\end{split}
\end{equation}
Since there can be some ambiguity in the term ``signal'', the terms used for the different contributions will be explained here. 
In essence, ``Higgs signal'' refers to the 1HSM contribution from $h_1$ and $h_2$ without their interference to the continuum QCD background. 
Corresponding example diagrams at the amplitude-squared level  
for the Higgs signal, the Higgs--QCD interference and the continuum background
are shown in figure~\ref{fig:lo-lo}.
We want to note that at LO, the interference effects only arise when the two incoming gluons produce the $t \bar t$ pair in a colour singlet configuration.

At the NLO level, corrections to the continuum background have been known for many years~\cite{Nason:1987xz,Beenakker:1988bq,Beenakker:1990maa,Mangano:1991jk} and also
corrections up to NNLO are available in the literature~\cite{Barnreuther:2012wtj,Czakon:2012zr,Czakon:2012pz,Czakon:2013goa,Czakon:2015owf,Czakon:2016ckf,Catani:2019hip,Catani:2019iny}, yielding perturbative uncertainties
of around $10(5)\%$ at NLO(NNLO), and slightly increasing in the tail of the $t \bar t$ invariant mass distribution. Corresponding corrections to the  interference
are only known in approximations, where the heavy-quark loop is integrated out~\cite{Bernreuther:2015fts,Bernreuther:2017yhg,BuarqueFranzosi:2017jrj}.\footnote{In ref.~\cite{Hespel:2016qaf} a corresponding NLO K-factor was constructed averaging the K-factors for the signal and the background.}
However, as $M_{h_2} > 2m_t$, where $m_t$ is the top-quark mass, this approximation is not expected to be valid for the process at hand. Thus, here we would like to go beyond the heavy-quark approximation providing (almost) exact NLO predictions for the signal process and the interference between the heavy Higgs signal and the continuum. The rest of this section presents the details of this NLO computation, where we focus the discussion on the interference. In our numerical results presented in section~\ref{sec:resultsstable} we include NLO corrections consistently for the signal, the continuum background and the interference. 

\begin{figure}[t]
	\centering
	\includegraphics[width=\diagwidthfraction\linewidth*\real{1.2}]{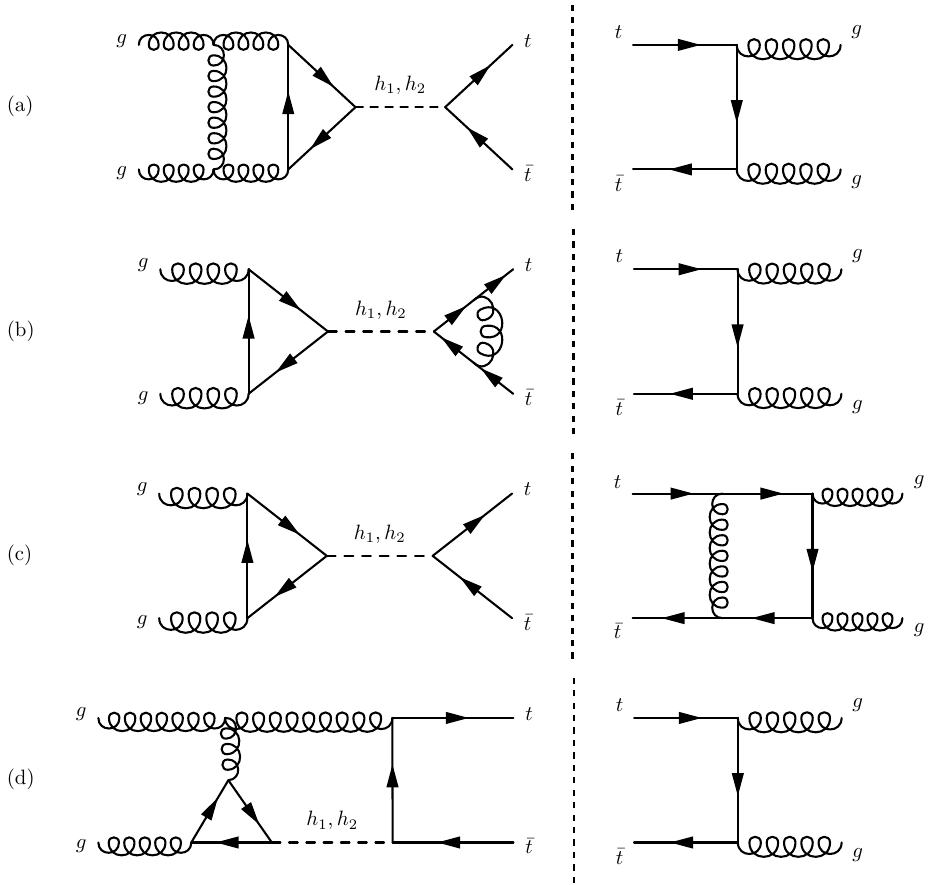}
	\caption{Example Feynman diagrams at the amplitude-squared level contributing to the virtual corrections to the interference at NLO. From top to bottom the diagrams represent corrections to Higgs production a), to the Higgs decay b), to the continuum c), and non-factorisable corrections d).}
	\label{fig:nlo}
\end{figure}

At NLO we have to consider virtual and real contributions. At the virtual level contributions can be categorised into corrections to Higgs production, corrections to the $h_{1}/h_{2} \to t \bar t$ decay, corrections to the background, and non-factorisable corrections. Representative diagrams for all four categories contributing to the NLO virtual corrections are shown in figure~\ref{fig:nlo}. Corrections to the continuum background as shown in figure~\ref{fig:nlo}(c) are of loop-squared type and are readily available in \textsc{OpenLoops} (see section \ref{sec:openloops}). Corrections to the production as shown in figure~\ref{fig:nlo}(a) are incorporated via form-factors interfaced to \textsc{OpenLoops} tree-level amplitudes. These two-loop form-factors have been extracted from the results in ref.~\cite{Harlander:2005rq}. Details are given in section~\ref{sec:ff}. Similarly, corrections to the decay as shown in figure~\ref{fig:nlo}(b) are evaluated via the same form-factor approach for the heavy-quark loop in the production stage at one-loop and an explicit loop computation in \textsc{OpenLoops} for the decay. Finally, the non-factorisable corrections as shown in figure~\ref{fig:nlo}(d) are beyond current loop technology and we compute these in the Higgs Effective Field Theory (HEFT) reweighted with LO mass effects, however considering only those contributions that yield IR divergent limits. Before discussing this latter approach in more detail below, we first have a look at the NLO real corrections.

\begin{figure}[htbp]
	\centering
	\includegraphics[width=\diagwidthfraction\linewidth*\real{1.1}]{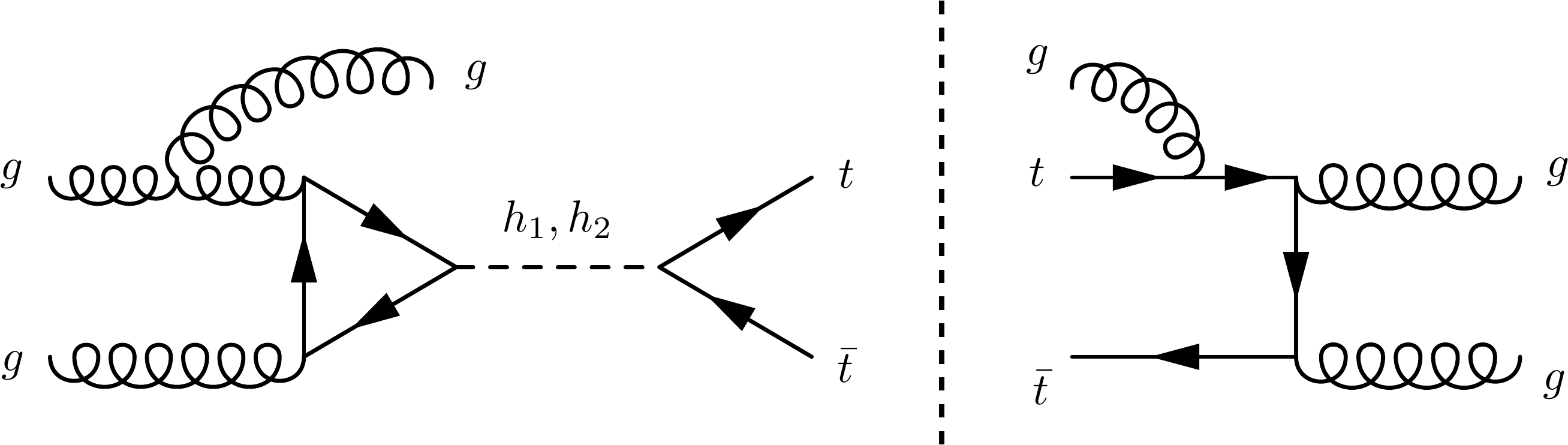}
	\caption{Example Feynman diagrams at the amplitude-squared level contributing to the real corrections to the interference at NLO. Shown is an infrared divergent non-factorisable contribution.}
	\label{fig:real-non-fact}
\end{figure}

\begin{figure}[htbp]
  \centering
  \hspace*{-6.3mm}\includegraphics[width=\diagwidthfraction\linewidth*\real{1.0}]{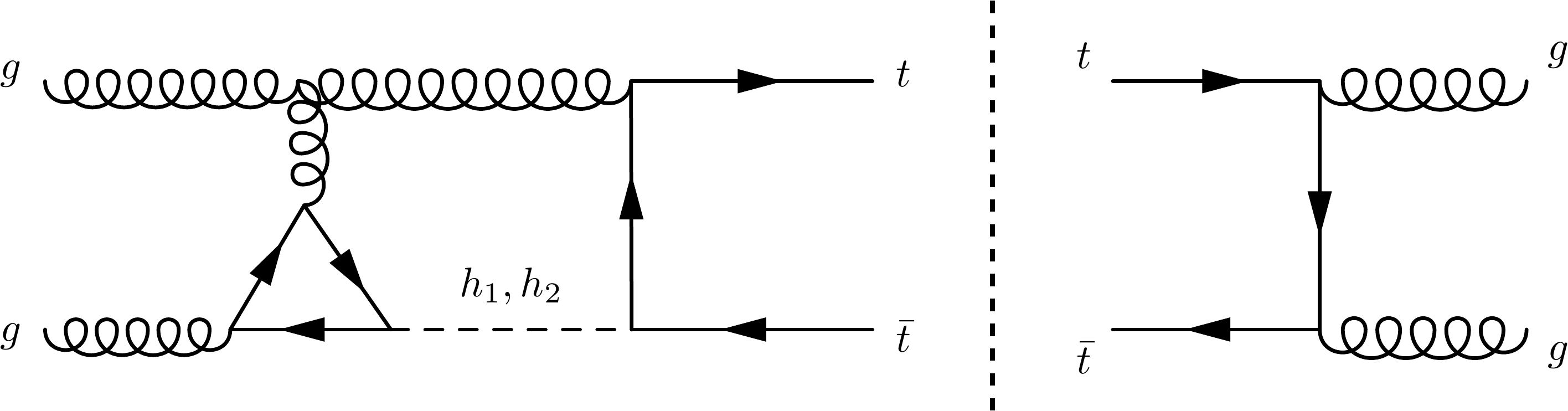}
  \includegraphics[width=\diagwidthfraction\linewidth*\real{1.0}]{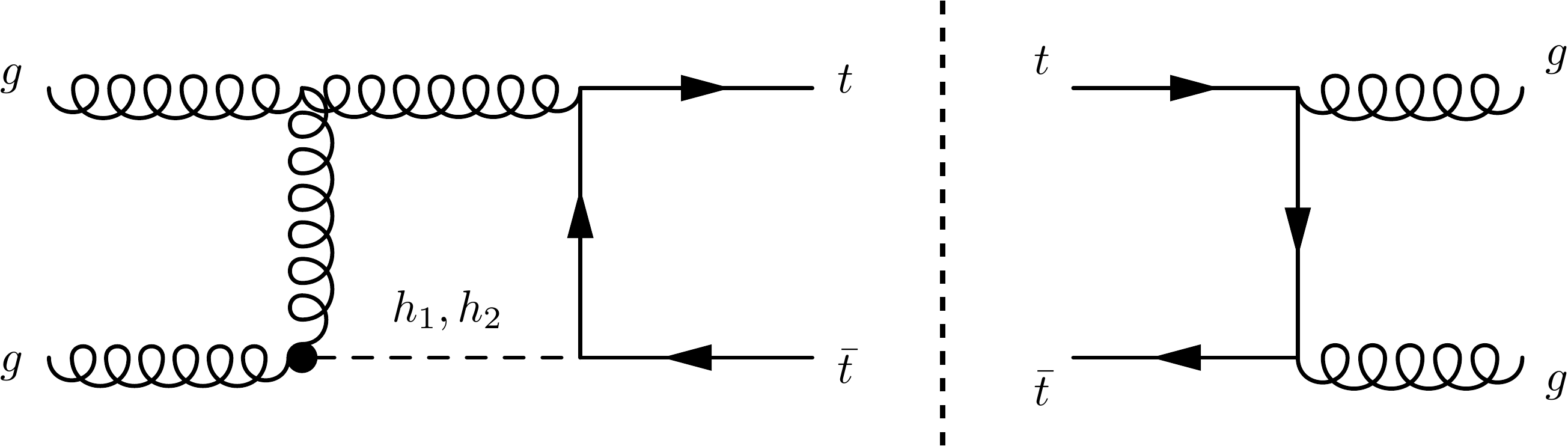}
  \caption{Example Feynman diagrams at the amplitude-squared level contributing to the virtual corrections to the interference at NLO. Shown is an infrared divergent non-factorisable contribution (top) together with the corresponding contribution in the HEFT (bottom).}
  \label{fig:virt-non-fact}
\end{figure}

\begin{figure}[htbp]
  \centering
  \includegraphics[width=\diagwidthfraction\linewidth*\real{1.0}]{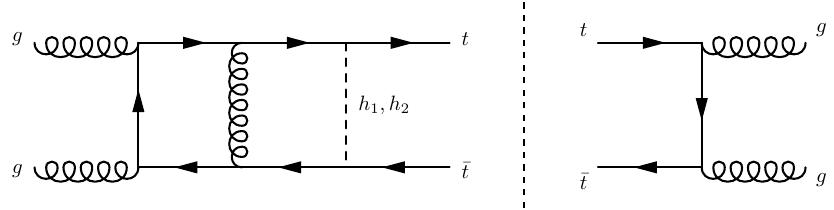}
  \caption{Example Feynman diagrams at the amplitude-squared level contributing to the virtual corrections to the interference at NLO. Shown is a finite non-factorisable contribution that cannot be treated in the approximation employed in our computation.}
  \label{fig:box-non-fact}
\end{figure}

For the computation of the NLO real corrections we have to evaluate 
all contributing amplitudes with an additional gluon, and the ones with an additional 
$q \bar q$ replacing a gluon at LO. In regard of real corrections to the Higgs--continuum interference, the amplitudes are given by born--loop interferences. Such amplitudes are readily available from automated one-loop amplitude providers like \textsc{OpenLoops}. However, these amplitudes have to be evaluated in infrared (IR) divergent limits, where the additional radiated parton can be soft or collinear to one of the LO initial-state gluons. Such phase-space regions pose severe challenges for the numerical stability of the one-loop amplitudes similar to real-virtual contributions in standard NNLO applications. We will discuss these issues in more detail in section~\ref{sec:openloops}.
Furthermore, these real radiation contributions to the interference generate non-factorisable corrections such as the
one displayed in figure~\ref{fig:real-non-fact}. Such contributions are non-zero for a (heavy) Higgs of
finite width, and do contain an infrared divergence. Correspondingly this infrared divergence needs to be cancelled by a corresponding virtual contribution.
In fact, the infrared divergence of the real correction of the type displayed in figure~\ref{fig:real-non-fact} is cancelled by non-factorisable virtual corrections such as the one displayed in figure~\ref{fig:virt-non-fact}.
However, the computation of such non-factorisable two-loop virtual
corrections is beyond today's loop technology, due to the presence of
three different masses in internal propagators.  Here we note that, in
the soft limit, these virtual contributions factorise into a singular
term times the interference between Higgs production at LO, with full
quark mass dependence in the loops, and the continuum production of
$t \bar t$ in an octet state. Therefore, the infrared divergent part
of the non-factorisable corrections can be reliably obtained by
computing all non-factorisable virtual corrections in the infinite
top-mass limit, dividing them by the LO contribution in the infinite
top-mass limit, and multiplying the result by the corresponding LO
contribution with full quark-mass dependence in the loops.  This is
the strategy we employ in this paper. We note that such rescaling
cannot account for non-factorisable corrections of the type displayed
in figure~\ref{fig:box-non-fact}, due to the absence of a closed top
quark loop. To summarise, after the rescaling procedure and the
cancellation of infrared divergences, the infrared-finite part of the
non-factorisable corrections arising from virtual diagrams will be
approximated, whereas all real corrections and integrated subtraction
counterterms will always be exact.

\begin{figure}[htbp]
  \centering
  \includegraphics[width=\diagwidthfraction\linewidth*\real{1.0}]{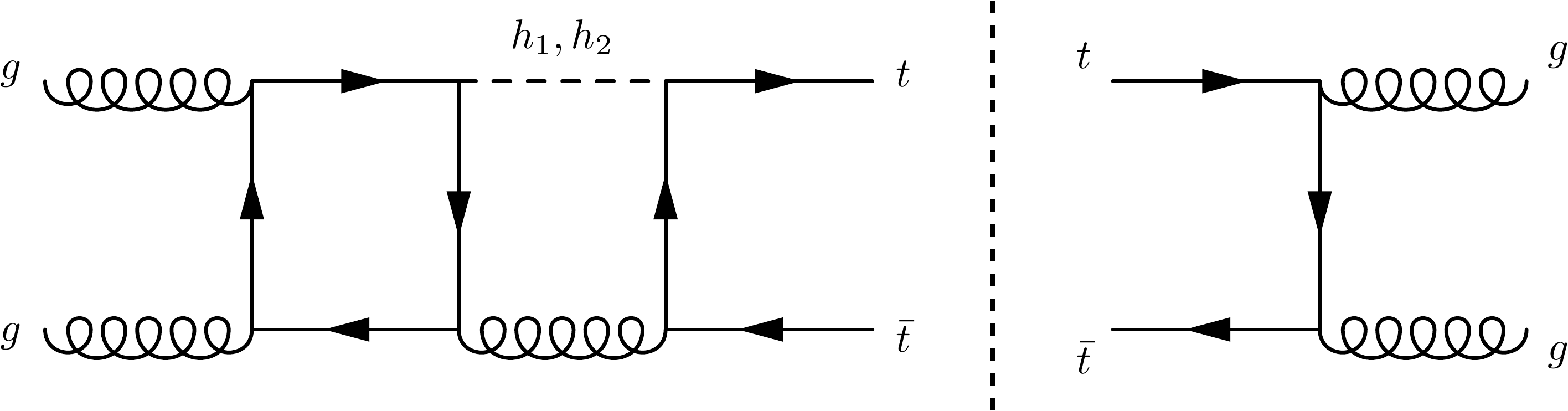}
  \includegraphics[width=\diagwidthfraction\linewidth*\real{1.0}]{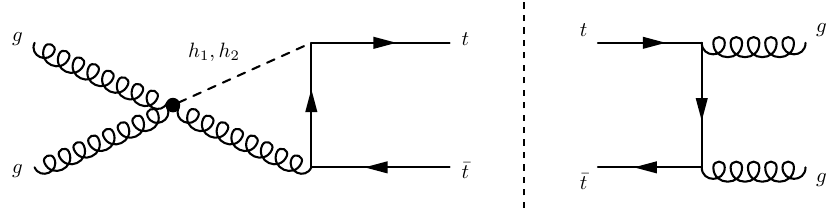}
  \caption{Example Feynman diagrams at the amplitude-squared level that is infrared finite and admits a HEFT counterpart. Shown is the contribution with the top quark loop (top) together with the corresponding contribution in the HEFT (bottom).}
  \label{fig:box-heftable}
\end{figure}

We observe that the so-computed virtual non-factorisable
  corrections are small, and correspond to $1\%$ of the total
  cross-section for all benchmark points. In order to have an
  estimate of the non-factorisable corrections that are not included
  in our approximation, we consider diagrams such as the one in
  figure~\ref{fig:box-heftable} that give rise to an effective
  three-gluon-Higgs vertex in the HEFT limit. We can then compute such
  contributions in the HEFT limit, and rescale them by the same factor
  as we did for the other non-factorisable corrections. The effect of
  these diagrams to the total cross section is less that $1\%$ for all benchmark points. Due to
  this, and the fact that these diagrams do not factorise from the
  Born amplitude in any limit, we decided not to include them in our
  nominal predictions.

Several of the ingredients discussed above are not readily available in public Monte Carlo tools. Therefore, we have compiled a new Monte Carlo framework, using \textsc{Kaleu}~\cite{vanHameren:2010gg} for phase space generation, and where the dipole subtraction \cite{Catani:1996vz, Catani:2002hc} has been derived from the corresponding implementation in \textsc{Helac-Dipoles}~\cite{Czakon:2009ss}. All tree and loop matrix elements have been obtained with a modified version of \textsc{OpenLoops}~2~\cite{Buccioni:2019sur}, where a dedicated interface to extract colour-correlated helicity amplitudes has been implemented. The two-loop virtual amplitudes have been incorporated via tree-level form-factors inserted into \textsc{OpenLoops} amplitudes, employing \textsc{CHAPLIN}~\cite{Buehler:2011ev} for the required harmonic polylogarithms~\cite{Remiddi:1999ew}. We have validated a resulting calculation of NLO QCD $pp \to t\bar t$
production against \textsc{MCFM}~\cite{Campbell:1999ah, Campbell:2011bn, Campbell:2015qma, Campbell:2019dru}, and that of Higgs production at NLO with
\textsc{SusHi}~\cite{Harlander:2012pb, Harlander:2016hcx}, finding perfect agreement in both cases. In the following, we refer to this computational setup as \textsc{Helac+OpenLoops}. 
Further details of this setup are given in section~\ref{sec:calcdetails}.


\section{Numerical setup}
\label{sec:calcdetails}

\subsection{Input parameters}
\label{sec:inputs}

For the numerical input parameters in our calculation we follow the recommendations from the LHC Higgs Working Group (LHCHWG)~\cite{LHCHiggsCrossSectionWorkingGroup:2016ypw}. The electroweak coupling $\alpha$ is determined in the $G_\mu$ scheme with,
\begin{equation}
	\alpha = \frac{\sqrt{2}}{\pi} G_{\mu} m_W^2 \sin^2 \theta_W \,, \hspace{3mm} \text{ and } \hspace{3mm} \sin^2\theta_W = 1-\frac{m_W^2}{m_Z^2} \,,
\end{equation}
where $G_{\mu}$ and the particle masses are given by
\begin{align}
	G_{\mu} &= 1.1663787 \times 10^{-5} \text{ GeV}^{-2} \,, & \nonumber \\
	m_W &= 80.35797 \text{ GeV} \nonumber \,, & 
	m_Z &= 91.15348 \text{ GeV} \nonumber \\
	m_t &= 173.2 \text{ GeV} \nonumber \,, & 
	m_b &= 4.92 \text{ GeV} \,, & \nonumber\\
	m_H &= 125 \text{ GeV} \,.  
\end{align}
We use the NLO PDF set \texttt{PDF4LHC15\_nlo\_mc} \cite{Butterworth:2015oua}, with $\alpha_s(M_Z) = 0.118$, as well as three-loop $\alpha_s$ running, for all cross sections, both LO and NLO. Factorisation and renormalisation scales $\mu_R, \mu_F$ are set  to
\begin{equation}
	\mu = \mu_R = \mu_F = \frac{M_{t\bar{t}}}{2} \,, \label{eq:scalecentralchoice}
\end{equation}
where $M_{t\bar{t}}$ is the invariant mass of the top-antitop pair.
The hadronic $pp$ collision energy is taken as $\sqrt{s} = 13$ TeV. 

\subsection{Higgs decay widths}
\label{sec:decaywidth}

\begin{table}[t]
\centering
\makebox[\textwidth][c]{
\begin{tabular}{lccccc}
\toprule
                            & $M_{h_2}$ {[}GeV{]}      & $700$                      & $1000$                     & $1500$                     & $3000$                     \\
\midrule
\multirow{4}{*}{$\theta_1$} & $\Gamma_{h_1}$ {[}GeV{]} & $3.910(5) \times 10^{-3}$  & $3.910(5) \times 10^{-3}$  & $4.004(5) \times 10^{-3}$  & $4.067(5) \times 10^{-3}$  \\
                            & $\Gamma_{h_1} / M_{h_1}$ & $3.1283(4) \times 10^{-5}$ & $3.1283(4) \times 10^{-5}$ & $3.2034(4) \times 10^{-5}$ & $3.2537(4) \times 10^{-5}$ \\
                            & $\Gamma_{h_2}$ {[}GeV{]} & $10.780(3)$                & $34.295(3)$                & $79.52(2)$                 & $86.70(3)$                 \\
                            & $\Gamma_{h_2} / M_{h_2}$ & $0.015400(4)$              & $0.034295(3)$              & $0.053013(7)$              & $0.028902(9)$              \\
\midrule
\multirow{4}{*}{$\theta_2$} & $\Gamma_{h_1}$ {[}GeV{]} & $3.488(5) \times 10^{-3}$  & $3.488(5) \times 10^{-3}$  & $3.813(5) \times 10^{-3}$  & $4.017(5) \times 10^{-3}$  \\
                            & $\Gamma_{h_1} / M_{h_1}$ & $2.7908(4) \times 10^{-5}$ & $2.7908(4) \times 10^{-5}$ & $3.0506(4) \times 10^{-5}$ & $3.2139(4) \times 10^{-5}$ \\
                            & $\Gamma_{h_2}$ {[}GeV{]} & $33.903(8)$                & $116.37(4)$                & $273.6(2)$                 & $322.5(2)$                 \\
                            & $\Gamma_{h_2} / M_{h_2}$ & $0.04843(2)$               & $0.11637(4)$               & $0.18240(8)$               & $0.10751(5)$               \\
\bottomrule
\end{tabular}}
\caption{Decay widths and $\Gamma/M$ ratios for the light and heavy Higgs bosons, $h_1$ and $h_2$, in the 1HSM extension for the considered benchmark points. The error is due to rounding and the numerical integration.}
\label{table:decaywidths}
\end{table}

We determine the widths of the light and heavy Higgs bosons in the 1HSM as 
\begin{align}
	\Gamma_{h_1} &= \Gamma_H\! \left( M_{h_1} \right ) \cos^2\theta \,, \\
	\Gamma_{h_2} &= \Gamma_H\! \left( M_{h_2} \right ) \sin^2\theta + \Gamma\!\left ( h_2 \rightarrow n \times h_1 \right ) \,,
\end{align}
where $\Gamma_H(M)$ refers to the decay width of a SM Higgs with mass $M$. We compute the SM Higgs decay width at NLO QCD and electroweak using \textsc{Prophecy4f} \cite{Bredenstein:2006rh, Bredenstein:2006nk, Bredenstein:2006ha} for the $WW$ and $ZZ$ decay channels, and \texttt{HDECAY} \cite{Djouadi:1997yw, Djouadi:2018xqq} for the remaining decay channels. The calculated width is $\Gamma_H(125 \, \text{GeV}) = 4.087(5) \times 10^{-3} \; \text{GeV}$ which agrees with the current LHCHWG recommendation of $4.088 \times 10^{-3} \; \text{GeV}$~\cite{LHCHiggsCrossSectionWorkingGroup:2016ypw}.
Due to numerical issues in \texttt{HDECAY} for large values of $M$ together with small values of $m_b$ for decay modes into $b$ quarks, we approximate $\Gamma_H \approx \Gamma(H \to WW) + \Gamma(H \to ZZ)$ for the benchmark points with $M_{h_2} = 3 \text{ TeV}$. Besides the SM contribution $\Gamma_H$ the decay width of the heavy Higgs  $\Gamma_{h_2}$ also receives contributions from $\Gamma\left ( h_2 \rightarrow n \times h_1 \right )$. We consider these explicitly for $2 \leq n \leq 4$, as detailed in appendix~\ref{app:higgsdecay}. The resulting numerical values for the decay widths $\Gamma_{h_{1/2}}$  in the eight considered 1HSM benchmark scenarios as defined in section~\ref{sec:model} are listed in table~\ref{table:decaywidths}
together with their respective $\Gamma/M$ ratios. The latter are crucial to gauge the size of non-factorisable corrections. As expected, they are very small for the SM Higgs, but get above 10\% for $M_{h_2}=1000,1500,3000\,$GeV and $\theta=\theta_2$.

\subsection{One-loop amplitudes and numerical stability}
\label{sec:openloops}

The LO and one-loop amplitudes implemented in \textsc{OpenLoops} have been compared at the amplitude level for several phase space points to the implementation used in ref.~\cite{Kauer:2019qei}, and were found to be in agreement. The implementation of the continuum QCD background for $pp \to t\bar{t}$ at NLO in \mcname{} has been validated against the result from \textsc{MCFM} \cite{Campbell:1999ah, Campbell:2011bn, Campbell:2015qma, Campbell:2019dru}, and was also found to be in agreement. The implementation of the dipole integrated subtraction terms in  \textsc{Helac-Dipoles} was validated against the corresponding implementation in \textsc{OpenLoops} for the continuum QCD background.

For numerical stability, a small cut on the transverse momentum of the real radiated jet is applied, $p_{T,j} > p_{T, \min}$. This is in particularly necessary for the evaluation of the loop-induced real emission amplitudes in the $gg$-channel, $gg \to Hg$, which can suffer from numerical instabilities in the penta-box diagrams. To produce all results in this paper we have set $p_{T, \min}=0.1\,$GeV. Additional stability treatment is included in \mcname{}, which discards events based on the relative accuracy estimate provided by \textsc{OpenLoops}, which will be above some threshold for events causing numerical instability in the loop calculations. A cut on the relative ratio of the invariant mass of the emitter, $s_{ij} / s_{ijk} < \epsilon$, has been applied. Theresults in this paper correspond to $\epsilon=10^{-6}$. Furthermore, events that are in the soft and collinear region, i.e.\ low $p_T$ or low $s_{ij} / s_{ijk}$ have been flagged for a further stability check of the relative error of the real-subtracted cross section. We have checked that the effect of varying the cutoffs $p_{T, \min}$ and $\epsilon$ around the values given above is within Monte Carlo error.

\subsection{One- and two-loop form factors for gluon-fusion Higgs production}
\label{sec:ff}

The one- and two-loop form factors for $gg \to H$ production \cite{Spira:1995rr, Harlander:2005rq} have been implemented in \textsc{OpenLoops} with finite top and bottom mass corrections. Exact mass dependence of the form factors are especially important for a BSM model with a light Higgs mass, $M_{h_1} < 2 m_t$, and a heavy Higgs mass, $M_{h_2} > 2 m_t$. The code for the form factors was adapted partly from \textsc{JetVHeto}~\cite{Banfi:2015pju} and partly from the \texttt{gg\_H\_quark-mass-effects} process \cite{Bagnaschi:2011tu} in \texttt{POWHEG BOX V2} \cite{Nason:2004rx, Frixione:2007vw, Alioli:2010xd}.

The form factor representation for the coupling of a Higgs doublet to two on-shell gluons of momenta $q_1$ and $q_2$, colour indexes $a$ and $b$, and polarisations $\varepsilon_{\pm}^\mu(q_1)$ and $\varepsilon_{\pm}^\nu(q_2)$ respectively~\cite{Davies:2019wmk} is given by
\begin{equation}
\mathcal{V}^{\mu\nu, a b}(q_1,q_2) = \frac{\alpha_s}{4 \pi v} F \,\delta^{ab} \left ( (q_1\cdot q_2)\, g^{\mu\nu} - q_{1}^{\nu}\, q_{2}^{\mu} \right ) \,,
\end{equation}
where the form factor $F$ can be represented as a series expansion in powers of $\alpha_s$,
\begin{equation}
	F = F_1 + \frac{\alpha_s}{2\pi} F_2 + \mathcal{O}(\alpha_s^2) \,. \label{eq:totalformfactor}
\end{equation}
The one-loop form factor $F_1$ is given by~\cite{Harlander:2005rq}
\begin{equation}
	F_{1} = -\sum_q \frac{2}{\tau_q^2} \left [ \tau_q + \frac 14(1-\tau_q) \ln^2 x_q \right ] \,,
\end{equation}
where
\begin{equation}
	\tau_q \equiv \frac{q_1 \cdot q_2}{2 m_q^2} \,, \hspace{15mm} x_q \equiv \frac{\sqrt{1 - \tau_q^{-1}} - 1}{\sqrt{1 - \tau_q^{-1}} + 1} \,.
\end{equation}
The index $q = b, t$ refers to the heavy-quark flavours in the loop,
with quark mass $m_q$. The
correct analytic continuation for all functions of $x_q$ can be
obtained via the replacement $\tau_q\to \tau_q+i 0$.

The two-loop form factor $F_2$, UV renormalised in the $\overline{\text{MS}}$ scheme for the strong coupling and the on-shell scheme for the quark masses $m_q$,  and in $4-2\epsilon$ dimensions, is given by
\begin{multline}
		F_{2} = \left(\frac{4\pi \mu_R^2}{-2(q_1\cdot q_2)-i0}\right)^\epsilon \frac{1}{\Gamma(1-\epsilon)}\left \{-\left(\frac{C_A}{\epsilon^2}+\frac{\beta_0}{\epsilon}+\beta_0\ln\left(\frac{2(q_1\cdot q_2)}{\mu_R^2}\right)\right) F_1 \right. \\ \left. +2 \sum_{q} \left[C_F \left(\mathcal{F}_{1/2}^{2l,a}(x_q) + \frac{4}{3} \mathcal{F}_{1/2}^{2l,b}(x_q) \right)+ C_A \mathcal{G}_{1/2}^{2l}(x_q) \right ]  \right\}\,,
\end{multline}
where $\mu_R$ is the renormalisation scale and $\beta_0=(11 C_A-2 n_f)/6$. The definitions of the functions $\mathcal{F}_{1/2}^{2l,a}(x), \mathcal{F}_{1/2}^{2l,b}(x)$ and $\mathcal{G}_{1/2}^{2l}(x)$ can be found in 
ref.~\cite{Aglietti:2006tp}.

The implementation of the one- and two-loop form factors for gluon-fusion Higgs production in \textsc{OpenLoops} has been validated by considering the process $pp \to H$ at NLO in \mcname{} and comparing the total cross section to the result from \textsc{SusHi}~\cite{Harlander:2012pb, Harlander:2016hcx}. The $pp \to H$ NLO cross section has been validated against \textsc{SusHi} for varying Higgs mass $m_H$ and scales, $\mu_R$ and $\mu_F$, to ensure the correct behaviour of the form factors. All cross sections were found to be in agreement with \textsc{SusHi} within their respective Monte Carlo error estimates. The $p_T$ distribution for $pp \to H$ from \mcname{} has been validated against \textsc{H1jet} \cite{Lind:2020gtc}.


\section{Results}
\label{sec:resultsstable}

In this section we present NLO predictions for the process $pp \; ( \to \{ h_1, h_2 \}) \to t\bar{t} + X$ in the SM and the 1HSM at the inclusive level, i.e.\ with stable tops,
and no fiducial cuts applied. Presented results are for the LHC with a centre-of-mass energy $\sqrt s =13\,$TeV. In this discussion we separate out various contributions to the $t \bar{t}$ cross-section, as defined in eq.~\eqref{eq:1hsmcontributions}.

\subsection{Inclusive cross sections}
\label{sec:resultsstablexsec}

We first consider the $pp \to t\bar{t}$ process in the SM, i.e.\ with $\mathcal{M}_{h_2}=0$. The various contributions to the  inclusive NLO cross-section and associated $K$-factors, $K =\sigma_{\text{NLO}}/\sigma_{\text{LO}}$, for the QCD background, the Higgs signal, and for the Higgs--QCD interference are shown in table~\ref{table:inclusiveresults} (upper part).
\begin{table}[tbh]
\centering
\bgroup
\def\arraystretch{1.2} 
\setlength\tabcolsep{11pt} 
\makebox[\textwidth][c]{
\begin{tabular}{cc|cc|cc}
\toprule
\multicolumn{6}{c}{$pp \; ( \to \{ h_1\}) \to t\bar{t} + X$ in the SM} \\
\midrule
 \multicolumn{2}{c}{QCD background} &\multicolumn{2}{c}{Higgs signal}            &     \multicolumn{2}{c}{Higgs-QCD Interference} \\
$\sigma_{\text{NLO}}^{\rm QCD}$ {[}pb{]} & $K^{\rm QCD}$        & $\sigma^{\rm Higgs}_{\text{NLO}}$ {[}pb{]} & $K^{\rm Higgs}$        & $\sigma^{\rm interf}_{\text{NLO}}$ {[}pb{]} & $K^{\rm interf}$        \\
\midrule
 $675.23(4)$                     & $1.5965(1)$ & $0.030971(3)$                     & $1.6512(2)$& $-1.56498(6)$                     & $2.1509(2)$ \\
\bottomrule
\\
\toprule
\multicolumn{6}{c}{$pp \; ( \to \{ h_1, h_2 \}) \to t\bar{t} + X$ in the 1HSM} \\
\midrule
                            &                     & \multicolumn{2}{c|}{Higgs signal}      & \multicolumn{2}{c}{Higgs--QCD interference}  \\
                            & $M_{h_2}$ {[}GeV{]} & $\sigma^{\rm Higgs}_{\text{NLO}}$ {[}pb{]} & $K^{\rm Higgs}$        & $\sigma^{\rm interf}_{\text{NLO}}$ {[}pb{]} & $K^{\rm interf}$     \\
\midrule
\multirow{4}{*}{$\theta_1$} & $700$               & $0.029108(2)$                     & $1.6234(2)$ & $-1.474(3)$                     & $2.112(4)$ \\
                            & $1000$              & $0.027334(2)$                     & $1.6459(2)$ & $-1.49140(8)$                    & $2.1584(2)$ \\
                            & $1500$              & $0.029932(3)$                     & $1.6745(2)$ & $-1.53751(8)$                     & $2.1609(2)$ \\
                            & $3000$              & $0.030933(3)$                     & $1.6661(2)$ & $-1.5800(2)$                     & $2.1822(2)$ \\
\midrule
\multirow{4}{*}{$\theta_2$} & $700$               & $0.027231(2)$                     & $1.5689(2)$ & $-1.289(3)$                     & $2.043(5)$ \\
                            & $1000$              & $0.020114(2)$                     & $1.6442(2)$ & $-1.30284(6)$                    & $2.1382(2)$ \\
                            & $1500$              & $0.026519(2)$                     & $1.6617(2)$ & $-1.44509(3)$                     & $2.13868(8)$ \\
                            & $3000$              & $0.029772(2)$                     & $1.6452(2)$ & $-1.53725(7)$                     & $2.1507(2)$ \\
\bottomrule
\end{tabular}}
\egroup
\caption{Integrated NLO cross-sections and corresponding $K$-factors for the LHC process $pp \, ( \to \{ h_1, h_2 \}) \to t\bar{t} + X$ with $\sqrt{s} = 13$ TeV in the SM (upper part) and the 1HSM (lower part) for each of the considered eight benchmark points. For the 1HSM results ``Higgs signal'' and ``Higgs-QCD interference''  refer  to the respective sums of the $h_1$ and $h_2$ contributions as defined in (\ref{eq:1hsmcontributions}). MC integration errors are shown in brackets.}
\label{table:inclusiveresults}
\end{table}
First, we note that at the fully inclusive level the Higgs signal is very small compared to the QCD background. This is expected due to the far off-shell SM Higgs propagator. The interference between the Higgs and the QCD background yields a negative subpercent contribution compared to the background.
In magnitude the interference is about 1.5 orders of magnitude larger than the Higgs signal. The NLO computation of this interference term in the SM and in the 1HSM with full mass dependence (apart from in the non-factorisable corrections, as discussed in section~\ref{sec:nlocorrinterf}) represents the main result of the paper at hand.
It is useful at this stage to compare our results to that of ref.~\cite{Hespel:2016qaf}. In their ansatz, the NLO interference is obtained by rescaling the LO interference with the geometric average of the $K$-factors for the Higgs signal and the QCD background,
\begin{equation}
	\sigma_{\rm NLO}^{\rm interf} = \sqrt{K^{\rm Higgs} \cdot K^{\rm QCD} } \, \sigma_{\rm LO}^{\rm interf}
\end{equation}
This ansatz yields $K^{\rm interf}_{\rm estimate}= \sqrt{K^{\rm Higgs} \cdot K^{\rm QCD} }=1.62$ in contrast to $K^{\rm interf}=2.01$ we obtain via the explicit computation, i.e.\ the ansatz of ref.~\cite{Hespel:2016qaf} predicts an interference at NLO of $-1.18\,$pb, while the explicit computation yields $-1.46\,$pb.

In table~\ref{table:inclusiveresults} (lower part), we show the contributions of the Higgs signal and of the Higgs--QCD interference to the NLO total cross section in the 1HSM for the benchmark points of table~\ref{table:1hsmmodel}.
At the inclusive level we observe no appreciable modifications of the NLO K-factors for any of the 1HSM cross-sections compared to the SM. 
In the next Section~\ref{sec:resultsstabledist} we move on to discuss 
corresponding NLO predictions at the differential level.

\subsection{Differential distributions}
\label{sec:resultsstabledist}

\begin{figure}[htbp]
	\centering
	\setlength{\abovecaptionskip}{0pt}
	\includegraphics[width=\linewidth]{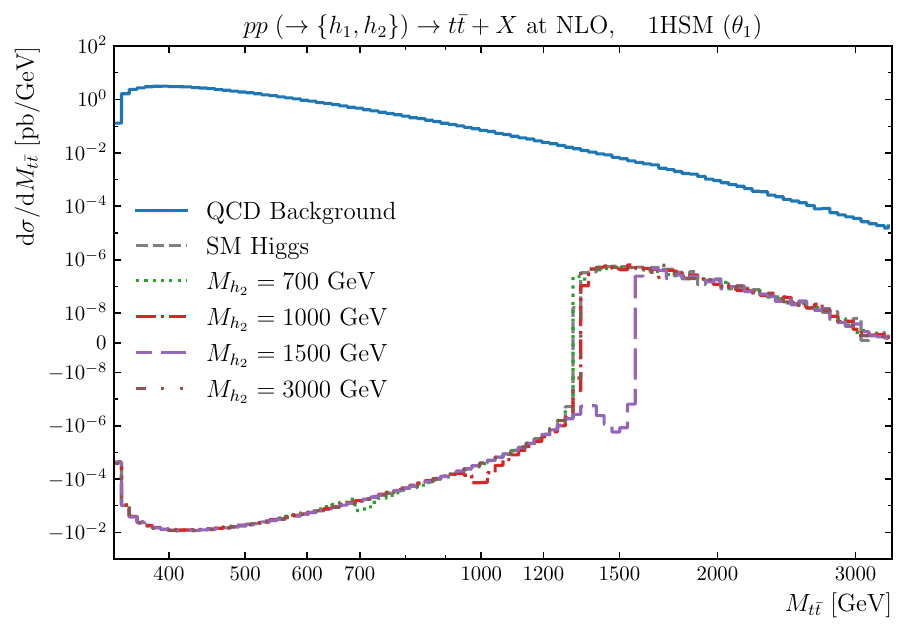}
		\includegraphics[width=\linewidth]{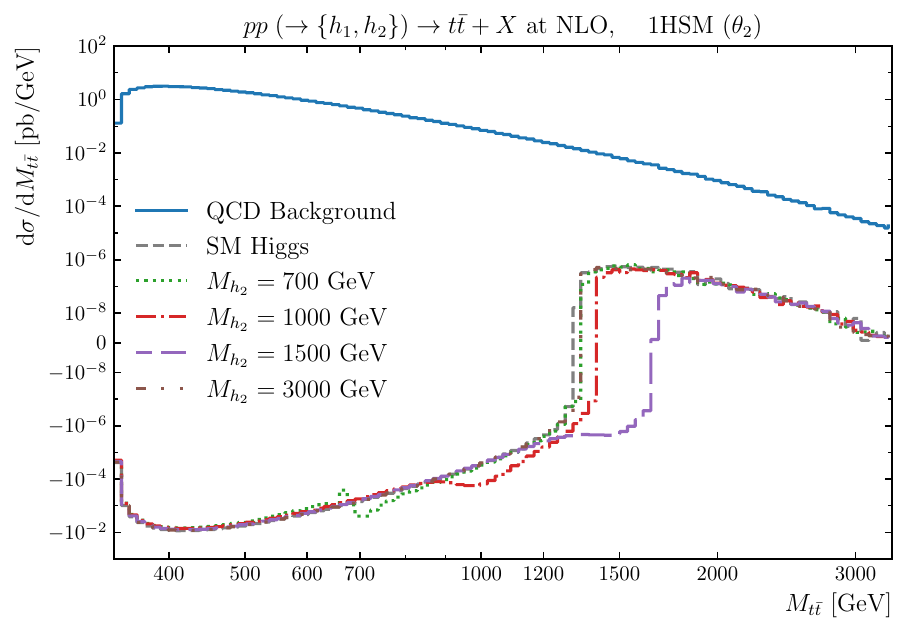}
	\captionof{figure}{NLO differential distribution in $M_{t\bar{t}}$ for the 
	QCD continuum $pp \to t\bar{t} + X$, the SM Higgs mediated process $pp \, ( \to h ) \to t\bar{t} + X$ and the BSM process $pp \, ( \to h_1,h_2 ) \to t\bar{t} + X$ at NLO in the 1HSM for several benchmark scenarios with $\theta = \theta_1$ (upper plot) and $\theta = \theta_1$ (lower plot). The ${\rm d}\sigma/{\rm d}M_{t\bar{t}}$ axis is plotted in log-scale with a linearised section between $-10^{-8}$ and $10^{-8}$.}
	\label{fig:mttbar1}
\end{figure}

We consider distributions in the invariant mass of the $t\bar{t}$ system, $M_{t\bar{t}}$, for the SM and the BSM benchmark scenarios of table~\ref{table:1hsmmodel}. In figure~\ref{fig:mttbar1}, we plot 
the corresponding NLO differential distributions for $\theta = \theta_1$ (upper plot) and $\theta = \theta_2$ (lower plot) over a large mass range. In each figure we separate the relevant contributions according to eq.~\eqref{eq:1hsmcontributions}, namely the QCD background ($\left \vert \mathcal{M}_{\text{QCD}} \right \vert^2$), the SM Higgs contributions ($\left \vert \mathcal{M}_{h_1} \right \vert^2 + 2 \, \mathrm{Re} ( \mathcal{M}_{h_1}^{\ast} \mathcal{M}_{\text{QCD}} )$) and the SM+BSM Higgs contributions ($\left \vert \mathcal{M}_{h_1} \right \vert^2 + \left \vert \mathcal{M}_{h_2} \right \vert^2 + 2 \, \mathrm{Re} \left ( \left ( \mathcal{M}_{h_1}^{\ast} + \mathcal{M}_{h_2}^{\ast} \right ) \mathcal{M}_{\text{QCD}} \right ) + 2 \, \mathrm{Re} ( \mathcal{M}_{h_1}^{\ast} \mathcal{M}_{h_2}^{\ast} )$). The latter are labelled according to the value of the mass of the heavy Higgs. 
The Higgs contributions are suppressed by about two orders of magnitude with respect to the QCD background. The BSM contributions largely follow the SM continuum apart from the peak/dip structure around the mass of the heavy Higgs. A precise modelling of these structures is crucial in order to separate the overwhelming QCD background. 
The SM Higgs and the BSM Higgs predictions are largely dominated by the interference contributions yielding negative cross-sections up to $M_{t\bar{t}} \approx 1300$ GeV. The SM Higgs contribution turns positive at around $M_{t\bar{t}} \approx 1300$ GeV yielding a 
spurious enhanced shape in the chosen linearised log plot for all SM Higgs and BSM Higgs predictions.
 We note that for $\theta = \theta_2$ the deviations around the mass of the heavy Higgs are smeared out compared to those for $\theta = \theta_1$. This is expected as the width of the resonances for $\theta = \theta_2$ are larger than for $\theta = \theta_1$, as listed in table~\ref{table:decaywidths}. The peak/dip structure for the BSM scenario with $M_{h_2}=3000$ GeV is marginally visible in these plots and we will neglect it in the further discussion. Such high masses of the heavy Higgs will be out-of-reach even at the HL-LHC.

\begin{figure}[htbp]
	\centering
	\setlength{\abovecaptionskip}{0pt}
	\includegraphics[width=\linewidth]{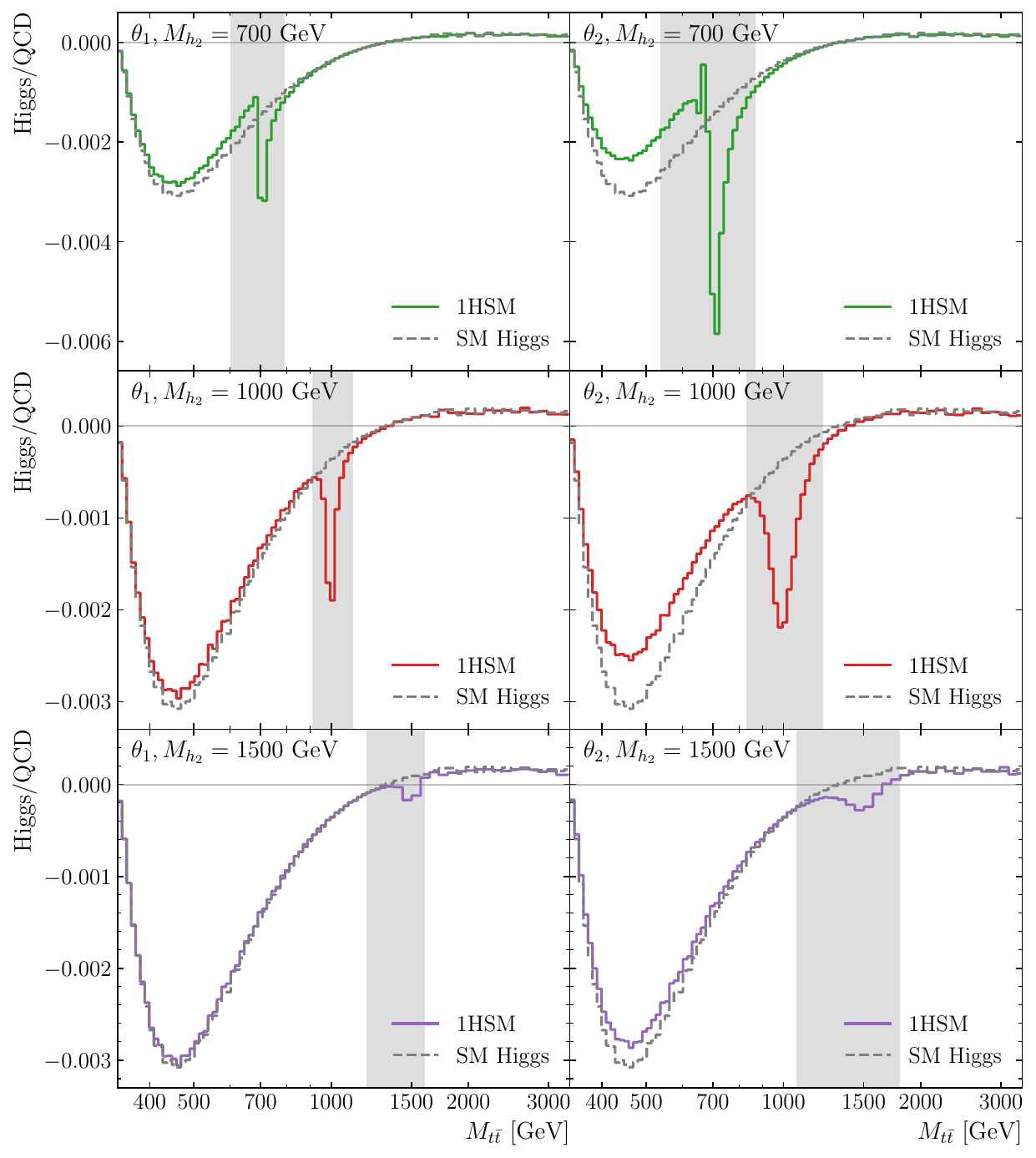}
	\captionof{figure}{Ratio of the Higgs contribution (signal and interference) for the process $pp \, ( \to \{ h_1, h_2 \} ) \to t\bar{t} + X$ to the continuum QCD background at NLO differential in the top-pair invariant mass, $M_{t\bar{t}}$, for each 1HSM benchmark point and in the SM. The chosen invariant mass windows are represented by a grey band. }
	\label{fig:ratio}
\end{figure}

\begin{figure}[htbp]
	\centering
	\setlength{\abovecaptionskip}{0pt}
	\includegraphics[width=\linewidth]{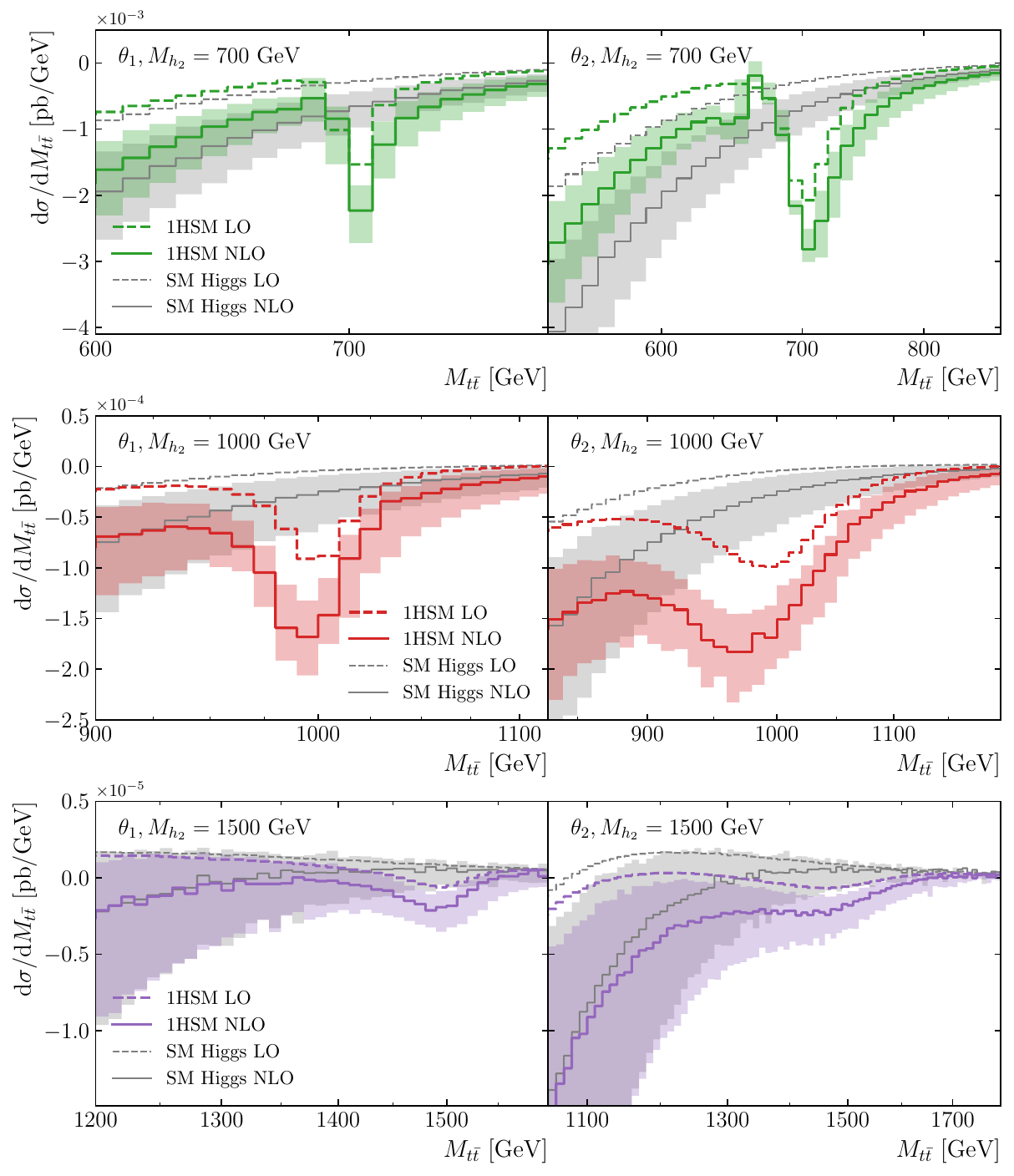}
	\captionof{figure}{Differential distributions in $M_{t\bar{t}}$ for the combined 
	Higgs contribution (signal and interference) in the 	1HSM and in the SM to the process $pp \, ( \to \{ h_1, h_2 \} ) \to t\bar{t} + X$ comparing NLO (solid) and LO (dashed) predictions for each benchmark point in the respective invariant mass window as defined in table \ref{table:masswindows}. }
	\label{fig:nlo_vs_lo}
\end{figure}

In order to highlight the peak/dip structure in the BSM scenarios in more detail, which is the key objective of the paper at hand, in figure~\ref{fig:ratio} we consider again the NLO differential distribution in $M_{t\bar{t}}$ showing the relative difference
between the (B)SM Higgs predictions in the 1HSM/SM and the QCD continuum, labelled $\rm{Higgs/QCD} - 1$. A corresponding plot differential in the average transverse momentum of the top and anti-top quark is shown in appendix~\ref{app:distributions}.
For all benchmark points the BSM effects manifest as a rather sharp dip around the heavy Higgs mass on top of the SM Higgs continuum. This dip is significantly wider in the case of $\theta=\theta_2$ due to the increased heavy Higgs widths.  Additionally, for 
$\theta=\theta_2$ there are significant $h_1$-$h_2$ and off-shell $h_2$-QCD interference effects which alter the shape of the Higgs continuum even away from the resonance region. At the negative maximum of the relative contribution of the Higgs continuum at around $M_{t\bar{t}}\approx 450$~GeV these interference effects yield a relative reduction of around $30\%$.
Based on the shapes of the dip structures around the resonance regions we select suitable invariant mass windows for each benchmark point around the resonance structures. These invariant mass windows are highlighted as grey bands in figure~\ref{fig:ratio} and listed in table~\ref{table:masswindows}.
Within these invariant windows in figure~\ref{fig:nlo_vs_lo} we zoom into the respective differential distributions in the top-pair invariant mass and show absolute NLO and LO predictions for the SM and BSM processes $pp \, ( \to h_1,h_2 ) \to t\bar{t} + X$ for all considered benchmark points. NLO corrections are large reaching more than 100\% and typically broaden the resonance dip towards larger invariant masses. The shown uncertainty bands at NLO are obtained from renormalisation and factorisation scale variations based on the envelope of a seven-point variation around the central scale choice in eq.~\eqref{eq:scalecentralchoice}, i.e. considering the combinations $(\mu_R, \mu_F) = (2 \mu, 2 \mu), (\mu / 2, \mu / 2), (2 \mu, \mu), \allowbreak (\mu / 2, \mu), (\mu, 2 \mu), (\mu, \mu / 2)$. Scale uncertainties are at the level of $20-30\%$ in the resonance regions.
We checked that these NLO uncertainty bands do not overlap with corresponding bands at LO (not shown). Due to the large higher-order corrections affecting the processes at hand NLO is the first perturbative order where a reliable estimate of theoretical uncertainties can be inferred from scale variation.  In all invariant mass windows NLO corrections and uncertainties at NLO increase towards smaller invariant masses. This can be understood from  a shift of events from the resonance region to invariant masses above and below the resonance, and the steepness of the corresponding differential distribution.  

Last, we comment on the effect of non-factorisable
  corrections in the chosen mass windows. Virtual non-factorisable
  corrections included in our nominal predictions gives a contribution
  of order $5\%$ for all benchmark points in the total cross sections in
  the corresponding mass windows. Further non-factorisable virtual
  corrections such as the one in figure~\ref{fig:box-heftable}, and
  that are not included in our nominal predictions, are evaluated
  according to the rescaling described in
  section~\ref{sec:nlocorrinterf}, and give an additional contribution
  of order $1\%$.

\subsection{Sensitivity estimates to BSM effects}
\label{sec:discussion}

\begin{table}[tbp]
	\centering
	\bgroup
	\def\arraystretch{1.2} 
	\setlength\tabcolsep{7pt} 
	\makebox[\textwidth][c]{
		\begin{tabular}{ll|c|cccc}
 &   &                 &                                  & \multicolumn{3}{c}{$\Delta B_\mathrm{exp} / B$}  \\
\cpartline{5-7}
 &   &                 &                                  & \multicolumn{3}{c}{$|S| / \sqrt{B}$}  \\
 &                    & invariant     &                   & \multicolumn{3}{c}{$(|S| / \sqrt{B})\sqrt{\textrm{BR}_{2\ell2\ell'}}$}  \\
\cline{5-7}
 & $M_{h_2}$ {[}GeV{]} & mass window  & $|S| / B$ & Run 2      & Run 3      & HL-LHC     \\
	\hline\hline
	\multirow{9}{*}{$\theta_1$} & \multirow{3}{*}{$700$}  & \multirow{3}{*}{$600$--$790$ GeV} & \multirow{3}{*}{0.00012(3)}& 0.047 & 0.047 & 0.023 \\
\cpartline{5-7}
                                 &   &                         &                                     & 0.46(9) & 0.7(2) & 2.1(5) \\
                                 &   &                         &                                     & 0.10(2) & 0.15(3) & 0.47(9) \\
                                    \cline{3-7}
	                            & \multirow{3}{*}{$1000$} & \multirow{3}{*}{$900$--$1115$ GeV}  & \multirow{3}{*}{0.000476(6)}
	                            & 0.069 & 0.069 & 0.029 \\
\cpartline{5-7}
                                    &                         &                                     && 0.691(7) & 1.02(1) & 3.21(4) \\
                                    &                         &                                     && 0.154(2) & 0.226(3) & 0.713(7) \\
                                    \cline{3-7}
	                            & \multirow{3}{*}{$1500$} & \multirow{3}{*}{$1200$--$1600$ GeV} & \multirow{3}{*}{$6.3(2) \times 10^{-5}$}                                    & 0.072 & 0.072 & 0.03 \\
\cpartline{5-7}
                                  &  &                         &                                     & 0.051(2) & 0.076(3) & 0.239(8) \\
                                  &  &                         &                                     & 0.0114(4) & 0.0168(6) & 0.053(2) \\
	\hline
	\multirow{3}{*}{$\theta_2$} & \multirow{3}{*}{$1500$} & \multirow{3}{*}{$1050$--$1800$ GeV} & \multirow{3}{*}{0.000108(2)}  & 0.059 & 0.059 & 0.025 \\
\cpartline{5-7}
                                  &  &                         &                                     & 0.129(2) & 0.190(3) & 0.601(9) \\
                                  &  &                         &                                     & 0.0287(5) & 0.0422(7) & 0.133(2) \\
	\end{tabular}
}
	\egroup
	\caption{$S/B$ ratios and significance estimates for the 1HSM benchmark points for LHC Run 2, LHC Run 3 and HL-LHC. Results using statistical Poisson uncertainties are shown for the total cross section and the $4e$+$4\mu$+$2e2\mu$ final state. In addition to $S/B$, the relative total experimental uncertainty of the background, $\Delta B_\mathrm{exp} / B$, is given, as estimated in ref.~\cite{Durieux:2022cvf}. We note that $\theta_2$ for the $M_{h_2}$ values of $700$ and $1000$ GeV is already excluded experimentally.}
        \label{table:masswindows}
\end{table}

Based on the obtained NLO cross sections for the 1HSM benchmarks we can estimate the sensitivity of current and future LHC runs in excluding these BSM scenarios.  
Within each invariant mass window as defined in table~\ref{table:masswindows}
we consider $|S| / \sqrt{B} = \sqrt{\mathcal{L}} \, |\sigma_S| / \sqrt{\sigma_B}$ as a naive estimate for the significance from Poisson statistics \cite{Barlow:1993}, where $\sigma_S$ and $\sigma_B$ are the cross sections integrated over the appropriate $M_{t\bar{t}}$ window, and $\mathcal{L}$ is the integrated luminosity of the considered LHC scenario.
In particular, we consider the full LHC Run 2 data sample corresponding to an integrated luminosity of $\mathcal{L} = 139 \text{ fb}^{-1}$~\cite{ATLAS:2019pzw}, the projected integrated luminosity for LHC Run 3 of $\mathcal{L} \approx 300 \text{ fb}^{-1}$, and of $\mathcal{L} \approx 3000 \text{ fb}^{-1}$ for the future high luminosity (HL-LHC) upgrade \cite{ZurbanoFernandez:2020cco}. As signal cross section $\sigma_S$ we consider the deviation of the 1HSM cross section from the SM cross section. The 1HSM contributes the entire Higgs contribution including $h_1$ and $h_2$ (and their interference) and the Higgs-continuum interference. The subtracted SM contribution includes the SM Higgs and its continuum interference.  We note that the continuum $t\bar{t}$ contribution cancels in the signal cross section.
In this context, we point out that neglecting the SM subtraction when computing the signal cross section would result in a significant overestimation of the considered significances, as suggested by figure~\ref{fig:ratio}. For $M_{h_2}=700$ GeV, the amplification exceeds a factor $10$.
As background $\sigma_B$ we consider the full SM cross section including the SM Higgs cross section, the QCD continuum cross section and the SM Higgs-continuum interference. The fiducial cross sections for the Higgs signals are mostly negative due to sizable Higgs-QCD interferences. Therefore we pragmatically consider the absolute value in the significance estimate. The results are shown in table~\ref{table:masswindows}.\footnote{Due to a log-based bin partitioning, the precise windows are given by [$601.9$, $791.8$], [$908.1$, $1115.5$], [$1194.6$, $1607.8$] and [$1065.6$, $1802.3$] GeV, respectively.}

Our first significance estimates are based on the assumption that all top decay modes can be captured experimentally and the true $M_{t\bar{t}}$ distribution extracted.  A more practicable assumption is to measure the $t\bar{t}$ cross section using the $e^-e^+$, $\mu^-\mu^+$ and $e^\mp\mu^\pm$ decay modes \cite{ATLAS:2023slx}.  We then need to include the corresponding branching ratio $\textrm{BR}_{2\ell2\ell'}$, where $\ell,\ell'\in \{e,\mu\}$.  We use the theoretical Born result of $\textrm{BR}_{2\ell2\ell'}=4/9^2\approx 4.9\%$.  These second significance estimates, calculated as
\begin{equation}
\frac{|S|}{\sqrt{B}}\,\sqrt{\textrm{BR}_{2\ell2\ell'}}\,,
\end{equation} 
are also displayed in table~\ref{table:masswindows}.

We consider a benchmark point to be excludable if the significance is
larger than $2$. Using the full $t\bar{t}$ data set, we find that purely
based on statistical uncertainties all benchmark points with
$M_{h_2} = 700$ GeV and $1000$ GeV could only barely be excluded at
  the HL-LHC. However, this can only be seen as a simplified, very
preliminary estimate, since we do not take into account the top
decays, phase space cuts, or experimental systematics. Even this
simplified significance estimate cannot exclude benchmark points with
$M_{h_2} = 1500$ GeV in any of the three scenarios.
When the data set is limited to the $4e$+$4\mu$+$2e2\mu$ final state, no benchmark point is projected to be excludable.

We note that experimental efficiencies $\varepsilon < 1$ may further reduce the measured $t\bar{t}$ cross sections, and theoretical uncertainties also need to be taken into account. In table~\ref{table:masswindows}, in addition to $S/B$ the relative total experimental uncertainty of the background is given, as estimated in table~5 of ref.~\cite{Durieux:2022cvf}. This comparison highlights the need for a significant reduction of the background uncertainties, before a realistic chance to exclude the 1HSM in this channel at the HL-LHC will materialise.

In a comprehensive, realistic analysis, the prospects of exclusion at the LHC hence appear slim. On the other hand, when top decays are included applying more sophisticated, differential methods (see figure~\ref{fig:ratio}) to enhance the signal over background ratio and possibly increasing the data set by including additional $W$ boson decay modes in the analysis may result in sizeable improvements. We are encouraged by the fact that the semi-leptonic decay modes are now also being exploited in experimental $t\bar{t}$ analyses~\cite{ATLAS:2024vxm}. The situation will evidently improve at planned future colliders like the FCC-hh.


\FloatBarrier

\section{Conclusion}
\label{sec:conclusion}

In this work we considered the 1-Higgs-singlet extension of the SM including an additional heavy Higgs that can decay into a $t \bar t$ pair. The production of such a particle interferes with QCD $t\bar t$ production, and it is possible that the interference pattern leads to appreciable modifications of physical cross sections, for instance in the differential distribution in $M_{t\bar t}$, the invariant mass of the $t \bar t$ pair. Extending the analysis of ref.~\cite{Kauer:2019qei}, we have computed the full set of NLO corrections to the production of a scalar decaying into a  $t \bar t$ pair, for the case of stable top quarks. The corrections entail at the virtual level two-loop amplitudes containing Higgs bosons in interference with the QCD background at tree-level and one-loop Higgs boson amplitudes in interference with the one-loop QCD background, and corresponding loop-squared amplitudes at the real radiation level. The two-loop amplitudes containing Higgs bosons can be separated into ``factorisable'' and ``non-factorisable''. Factorisable two-loop diagrams are computed exactly. Some non-factorisable two-loop contributions are beyond today's loop technology. We hence include the non-factorisable two-loop contributions in an approximation which preserves all singular limits. The ingredients of our calculation are not available in current automated tools. Therefore, we built a custom Monte Carlo framework by acquiring amplitudes from a modified \textsc{OpenLoops} 2, using the dipole subtraction implementation from \textsc{Helac-Dipoles}, and \textsc{Kaleu} for phase-space integration. In this framework factorisable two-loop corrections are included via appropriate form-factors inserted into \textsc{OpenLoops} tree amplitudes.

In our numerical analysis we found that the presence of an additional Higgs boson induces distinct features in the $M_{t\bar t}$ distribution, located around the mass of the heavy Higgs. These features are affected by large NLO corrections, which broaden the $M_{t\bar t}$ distribution relative to the LO one. Performing a scan in $M_{t\bar t}$, we found that excluding the considered BSM model with the HL-LHC or a future collider like the FCC-hh will require differential discriminants, enlarged data sets and significant improvements in reducing the background uncertainties. An important step will be to upgrade our calculation by including the top decays, so that realistic experimental cuts can be applied and differential observables can be analysed. It would then be desirable to identify alternative observables to $M_{t\bar t}$ that show the same degree of sensitivity to BSM effects, but are less affected by real QCD radiation. We leave both tasks to future work. In summary, this study provides  the key ingredients to assess the impact of interference effects in BSM searches at NLO accuracy. The developed \textsc{Helac+OpenLoop} Monte Carlo program for the computation of $pp \; ( \to \{ h_1, h_2 \}) \to t\bar{t} + X$ in the SM and the 1HSM is available upon request.


\acknowledgments

A.L.\ is supported by the CNRS/IN2P3.  A.B., N.K.\ and J.L.\ are supported by the STFC under the Consolidated Grant ST/T00102X/1 and J.L.\ as well by the STFC Ernest Rutherford Fellowship ST/S005048/1. N.K.\ thanks the Technical University of Munich for hospitality. R.W. acknowledges the hospitality of the CERN Theory Group while part of this work was performed, and A.B.\ the hospitality of Royal Holloway University of London. This work was performed using the DiRAC Data Intensive service at Leicester, operated by the University of Leicester IT Services, which forms part of the STFC DiRAC HPC Facility (\href{https://dirac.ac.uk/}{www.dirac.ac.uk}). The equipment was funded by BEIS capital funding via STFC capital grants ST/K000373/1 and ST/R002363/1 and STFC DiRAC Operations grant ST/R001014/1. DiRAC is part of the UK National e-Infrastructure.


\appendix

\section{Decays of the heavy Higgs in the 1HSM}
\label{app:higgsdecay}

\begin{table}[tbp]
\centering
\makebox[\textwidth][c]{
\begin{tabular}{lccc}
\toprule
\multicolumn{4}{c}{$\theta = \theta_1$}                                                                                                                                                           \\
$M_{h_2}$ {[}GeV{]} & $\Gamma\left ( h_2 \to 2 \times h_1 \right )$ {[}GeV{]} & $\Gamma\left ( h_2 \to 3 \times h_1 \right )$ {[}GeV{]} & $\Gamma\left ( h_2 \to 4 \times h_1 \right )$ {[}GeV{]} \\
\midrule
$700$               & $2.1556(1)$                                             & $0.00468(2)$                                            & $6.24(4) \times 10^{-7}$                                \\
$1000$              & $6.0953(1)$                                             & $0.1692(7)$                                             & $0.001718(9)$                                           \\
$1500$              & $9.8911(1)$                                             & $0.218(2)$                                              & $0.001632(8)$                                           \\
$3000$              & $20.658(1)$                                             & $0.306(2)$                                              & $0.001060(7)$                                           \\
\bottomrule
                    &                                                         &                                                         &                                                         \\
\toprule
\multicolumn{4}{c}{$\theta = \theta_2$}                                                                                                                                                           \\
$M_{h_2}$ {[}GeV{]} & $\Gamma\left ( h_2 \to 2 \times h_1 \right )$ {[}GeV{]} & $\Gamma\left ( h_2 \to 3 \times h_1 \right )$ {[}GeV{]} & $\Gamma\left ( h_2 \to 4 \times h_1 \right )$ {[}GeV{]} \\

$700$               & $4.1798(1)$                                             & $0.507(2)$                                              & $0.01451(8)$                                            \\
$1000$              & $11.604(1)$                                             & $7.34(4)$                                               & $2.46(2)$                                               \\
$1500$              & $27.26(1)$                                              & $12.9(2)$                                               & $3.91(2)$                                               \\
$3000$              & $66.8(1)$                                               & $21.4(2)$                                               & $4.17(2)$                                               \\
\bottomrule
\end{tabular}}
\caption{Partial decay widths for $h_2 \to n \times h_1$ for $n = 2, 3,$ and $4$ in the 1HSM for the eight different benchmark points.}
\label{table:hdecaystheta}
\end{table}

\begin{figure}[h!]
	\centering
	\includegraphics[width=0.36\linewidth]{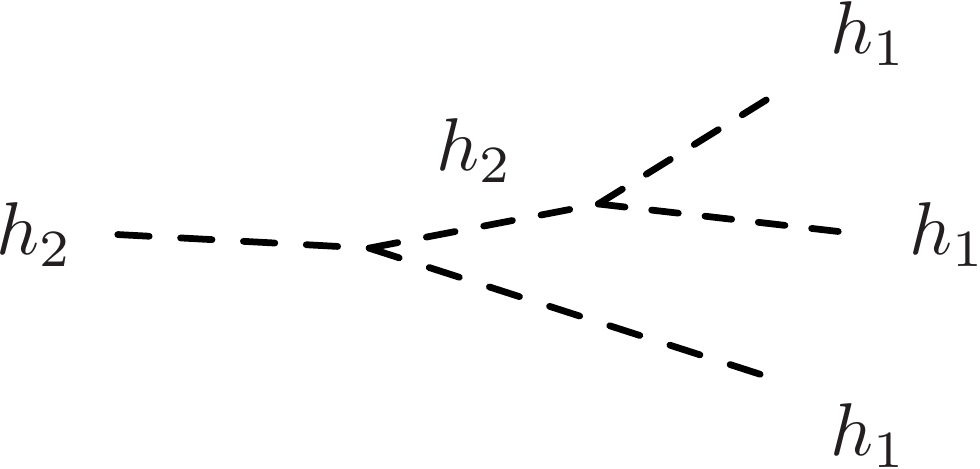}
	\caption{The cascaded partial decay of $h_2 \to 3 \times h_1$.}
	\label{fig:cascadeddecay}
\end{figure}

As detailed in section~\ref{sec:decaywidth} we determine the widths of the light and heavy Higgs bosons in the 1HSM as
\begin{align}
	\Gamma_{h_1} &= \Gamma_H\!\left( M_{h_1} \right ) \cos^2\theta \,, \\
	\Gamma_{h_2} &= \Gamma_H\!\left( M_{h_2} \right ) \sin^2\theta + \Gamma\!\left ( h_2 \rightarrow n \times h_1 \right ) \,,
\end{align}
where  $\Gamma_H(M)$ refers to the decay width of a SM Higgs with mass $M$, and $\Gamma\left ( h_2 \rightarrow n \times h_1 \right )$ to the
partial decay widths for the heavy scalar $h_2$ decaying into $n$ light scalars. 
We compute and include this latter partial decay width for $2 \leq n \leq 4$. A custom implementation of the 1HSM in \textsc{FeynRules} \cite{Alloul:2013bka, Christensen:2008py} was used to produce a UFO \cite{Degrande:2011ua} implementation that was subsequently used in \textsc{MadGraph5\_aMC@NLO} \cite{Alwall:2014hca} to calculate the partial decay widths $\Gamma\left ( h_2 \rightarrow n \times h_1 \right )$, the results of which are shown in table~\ref{table:hdecaystheta}. It is clear that partial decay widths for decays to higher multiplicities, $n > 4$, of $h_1$ are suppressed.
The partial decay widths for $n > 2$ is due to cascaded emissions of $h_2 \to h_2 h_1$ and decays $h_2 \to h_1 h_1$, as illustrated in figure~\ref{fig:cascadeddecay}.

The computation of the decay width $\Gamma_{h_2}$ implies a circular dependence due to the regularisation of the $h_2$ propagator via $\Gamma_{h_2}$ itself. This is not problematic if $\Gamma_{h_{2}} / M_{h_{2}} \ll 1$, but as can be seen in table~\ref{table:decaywidths} one of the benchmark points reaches $\Gamma_{h_{2}} / M_{h_{2}} \sim 0.18$. A simple solution would be to recursively calculate the decay width. However, in the case of $\Gamma/M \sim 1$ the Breit-Wigner approximation does not hold, and one should use a better estimate of the self energy $\Sigma(p^2)$ in the exact propagator expression. It is important, however, to do this in a gauge-invariant way. This would likely produce a tiny correction in this case, and has therefore not been done in this study. But it may be necessary for certain new physics models that introduces resonances with large decay widths.


\section{Additional differential distributions}
\label{app:distributions}

\begin{figure}[htbp]
	\centering
	\setlength{\abovecaptionskip}{0pt}
	\includegraphics[width=\linewidth]{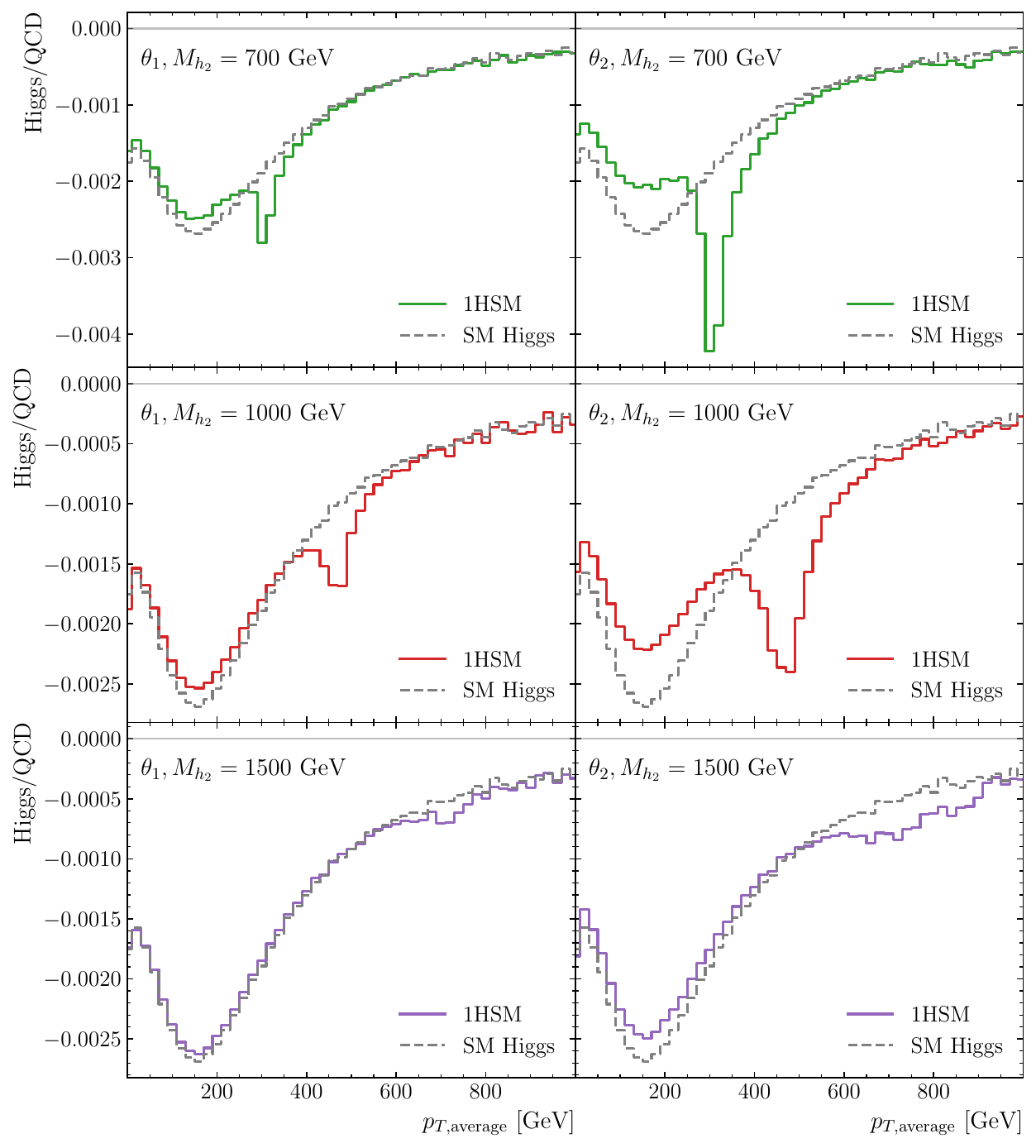}
	\captionof{figure}{Ratio of the Higgs contribution (signal and interference) for the process $pp \, ( \to \{ h_1, h_2 \} ) \to t\bar{t} + X$ to the continuum QCD background at NLO differential in $p_{T,\text{average}}$, for each 1HSM benchmark point and in the SM.  }
	\label{fig:ratiopTavg}
\end{figure}

In this appendix we include additional differential distributions for the process $pp \, ( \to h_1,h_2 ) \to t\bar{t} + X$ in the 1HSM. In figure~\ref{fig:ratiopTavg} we show differential distributions in the average $p_T$ of the $t$ and $\bar t$, $p_{T,\text{average}} = (p_{T,t} + p_{T,\bar{t}}) / 2$, for each 1HSM benchmark point normalised to the QCD continuum.

\FloatBarrier
\bibliographystyle{JHEP}
\bibliography{heavyhiggs.bib}


\end{document}